\documentclass[aps,showpacs,twocolumn]{revtex4-2}

\usepackage{hyperref}
\hypersetup{
    colorlinks=true,
    linkcolor=black,
    filecolor=black,      
    urlcolor=black,
    citecolor=black,
}

\usepackage{exscale}
\usepackage{graphicx}
\usepackage{amsmath}
\usepackage{latexsym}
\usepackage{epstopdf}
\usepackage{amsfonts}
\usepackage{amssymb}
\usepackage{bbm}
\usepackage{times}
\usepackage[T1]{fontenc}
\usepackage{lipsum}
\usepackage{amsthm}
\usepackage{fancyhdr,txfonts,bbm}
\usepackage{appendix}
\usepackage{soul}

\usepackage{epsfig,pstricks}
\usepackage[commandnameprefix=ifneeded]{changes}

\newcommand{{\Cd}}{{\mathbb{C}^d}}
\newcommand{{\C}}{{\mathbb{C}}}

\newcommand{\sign}{\text{sign}}

\newcommand{\expp}[1]{e^{#1}}
\newcommand{\exppp}[1]{\mbox{exp} \left[ #1 \right]}

\usepackage{overpic}

\begin{document}

\title{Ancillary Gaussian modes activate the potential to witness non-Markovianity}

\author{Dario De Santis$^1$, Donato Farina$^1$, Mohammad Mehboudi$^2$, and Antonio Ac{\'i}n$^{1,3}$}
\affiliation{$^1$ICFO-Institut de Ciencies Fotoniques, The Barcelona Institute of Science and Technology, 08860 Castelldefels (Barcelona), Spain \\ 
$^2$ Département de Physique Appliquée, Université de Genève, 1211 Genève, Switzerland\\
$^3$ ICREA - Instituci\'{o} Catalana de Recerca i Estudis Avan\c{c}ats, 08010 Barcelona, Spain
}

\date{\today}
\begin{abstract}

We study how the number of employed modes impacts the ability to witness non-Markovian evolutions via correlation backflows in continuous-variable quantum dynamics. We first prove the existence of non-Markovian Gaussian evolutions that do not show any revivals in the correlations between the mode evolving through the dynamics and a single ancillary mode. We then demonstrate how this scenario radically changes when two ancillary modes are considered. 
Indeed, we show that the same evolutions can show correlation backflows along a specific bipartition when three-mode states are employed, and where only one mode is subjected to the evolution. These results can be interpreted as a form of activation phenomenon in non-Markovianity detection and are proven for two types of correlations, entanglement and steering, and two classes of Gaussian evolutions, a classical  noise model and the quantum Brownian motion model.

\end{abstract}

\maketitle

\section{Introduction}

 The interaction between any given quantum system and the surrounding environment can never be completely avoided; the theory of open quantum systems is an indispensable framework to describe the realistic dynamics of quantum systems~\cite{book_B&P,book_R&H}.
 The interaction with the environment is usually detrimental for quantum resources, like quantum coherence or entanglement, and in general makes the resulting map on the system no longer unitary, but by a quantum channel, described by a completely-positive trace-preserving (CPTP) map.
The evolution in time is then described by a continuous family of quantum channels, which can be classified as Markovian or non-Markovian. While the former class is characterized by the continuous degradation of any type of information encoded in the system, in the latter the decoherence process is not monotonic in time---these \textit{recoherences} are often called backflows of information. Non-Markovian evolutions have attracted much interest, not only because of their fundamental interest, but also because the associated backflows can have a positive effect in various quantum information tasks, such as metrology \cite{metrology}, quantum key distribution  \cite{QKD}, quantum teleportation \cite{teleportation}, entanglement generation \cite{egeneration}, quantum communication \cite{Bognachannel}, information screening \cite{screening} and quantum thermodynamics \cite{T1,T2,T3,T4}. 
 
 The mathematical property used to define Markovianity is called CP-divisibility; the dynamics is Markovian if and only if it is possible to describe the evolution between any two times through the action of a physical quantum channel, that is, a CPTP map (for reviews on this topic see Refs.  \cite{rev_RHP,rev_Breuer,revmod2}).
Various strategies have been adopted to 
connect this mathematical definition to more physically motivated ones. This has lead to the development of witnesses of non-Markovianity through backflows of different quantities, such as the error probability in state discrimination \cite{BLP,BognaPRL,BD}, channel capacity \cite{Bognachannel}, Fisher information \cite{CinFish,AbiusoFish}, the volume of accessible states \cite{LPP} and correlations \cite{RHP,LFS,DDSshort,DDSMJ,Janek,ManiscalcoCV,ParisCV}.
In the case of correlations, the standard method for witnessing non-Markovianity works as follows: (i) prepare an initial state of two particles, which we name system and ancilla; (ii) apply the considered evolution to one of the two particles, the system, while the ancilla remains untouched and (iii) monitor how the correlations between the two particles change during the evolution. If the correlations do not decrease monotonically with time, the dynamics gives rise to a correlation backflow and therefore is non-Markovian.
Beyond this recipe,  in general it is not known whether and how to construct the initial two-particle state for a given non-Markovian dynamics or, even simpler, what is the minimal dimension of the ancillary system that is needed for this task. Even less is known for continuous-variable systems and, in particular, for Gaussian dynamics, despite their prominent role in many physically relevant scenarios.
While for finite-dimensional, several works have studied correlation backflows in quantum evolutions~\cite{desantis2020correlations,DDSMJ,Janek}, continuous-variable settings have not been explored beyond the use of a single ancillary mode \cite{ManiscalcoCV,ParisCV,GaussianInterferometricPowerNM}.

In this work, we firstly ask ourselves whether considering a single ancillary mode is sufficient to witness quantum correlation backflows for arbitrary non-Markovian dynamics. 
Particularly, we use both entanglement and Gaussian steerability as correlations. Our results show that indeed using a single ancillary mode is not always sufficient to witness backflows. 
Secondly, and motivated by this shortcoming, we  ask ourselves whether deploying a secondary ancillary system would  be advantageous.
 We show through two examples, namely  the dynamics of a single mode under (i) a classical noise model and (ii) the quantum Brownian motion model, that the secondary ancillary mode  allows witnessing non-Markovian evolutions that are impossible to detect with any possible single ancillary mode initialization. 
 Finally, we show that, while for some open dynamics two ancillary modes are sufficient for witnessing non-Markovianity, for some other dynamics one may need even a higher number of ancillary modes.
 
The article is structured as follows. 
In Sec.~\ref{sec:preliminaries} we briefly introduce Gaussian states and the entangled initializations of interest.
Such initializations are assumed to undergo local Gaussian dynamics, characterized in Sec.~\ref{gevo}, and their quantum correlations are quantified through Gaussian steerability and entanglement, Sec.~\ref{sec:NMwitnesses}.
The advantage stemming from the use of more than one ancillary mode is presented in Sec.~\ref{sec:examples}, through paradigmatic examples. 
Finally, we summarize our results in Sec.~\ref{sec:dicussion-conclusions}.
\section{Preliminaries}
\label{sec:preliminaries}
In this Section we set the notation and introduce the adopted formalism to describe quantum Gaussian systems. 
An $n$-mode continuous variable quantum system is defined through states over the Hilbert space $\mathcal{H}^{(n)}=\otimes_{i=1}^n \mathcal{H}_i$, 
where $\mathcal{H}_i$ is the Hilbert space of a bosonic harmonic oscillator corresponding to the $i$-th mode of the system. 
We call $S(\mathcal{H}^{(n)})$ the state space of density operators $\hat\rho$ associated to $\mathcal{H}^{(n)}$. 
The quadrature operators of the $i$-th mode are $\hat q_i=(\hat{a}_i+\hat{a}^\dagger_i)$ and $\hat p_i=-i(\hat{a}_i-\hat{a}^\dagger_i)$, where $\hat a_i$ ($\hat a^\dagger_i$) is the annihilation (creation) operator for the $i$-th mode. 
By grouping these operators in the vector $\hat X=(\hat q_1, \hat p_1, \hat q_2, \hat p_2, \dots )$, we can write the canonical commutation relations as $[\hat X_i, \hat X_j]= 2i \Omega^{(n)}_{ij}$, where
\begin{equation}
\Omega^{(n)}=\bigoplus_{i=1}^n \Omega^{(1)} \, , \hspace{0.75cm} \Omega^{(1)} = \left( \begin{array}{ccc} 0 & 1 \\ -1 & 0 \end{array} \right) ,
\end{equation}
the $2n\times 2n$ matrix $\Omega^{(n)}$ being the $n$-mode symplectic form and $\Omega^{(1)}$ the corresponding single-mode form.

A quantum state $\hat\rho\in S(\mathcal{H}^{(n)})$ is called \textit{Gaussian} when the first and second moment of the quadrature vector $\hat X$, namely 
\begin{equation}
d_i = \langle \hat X_i \rangle_{\hat\rho} \,\,\,\,\mbox{ and }\,\,\,\, \sigma_{ij}=\frac{1}{2} \langle\{\hat X_i, \hat X_j\} \rangle_{\hat\rho} - \langle \hat X_i \rangle_{\hat\rho} \langle \hat X_j \rangle_{\hat\rho} \, ,
\end{equation}
are sufficient to fully describe $\hat\rho$, where $\langle  \hat O \rangle_{\hat\rho}=\mbox{Tr}[ \hat\rho \hat O]$ is the expectation value of the operator $\hat O$ on the state $\hat\rho$. 
The $2n\times 2n$ real symmetric matrix $\sigma$ is called the \textit{covariance matrix} of the system. Two Gaussian states with different first moments and same covariance matrix can be mapped one into the other by a displacement unitary transformation. In the following we are interested on the information contained in the covariance matrix only and therefore we ignore $d_i$.

In case of a bipartite scenario, where Alice owns the first $n_A$ modes and Bob owns the last $n_B$, the covariance matrix $\sigma_{AB}$ of a shared Gaussian state can be written as follows: 
\begin{equation}\label{sigmaAB}
\sigma_{AB}= \left(
\begin{array}{ccc}
A  & C \\
{C}^T &  B
\end{array}
\right) \, ,\end{equation}
where the $2n_A\times 2n_A$ matrix $  A$ ($2n_B\times 2n_B$ matrix $  B$) is the covariance matrix of Alice's (Bob's) system and the correlation matrix $  C$ is $2n_A\times 2n_B$. In order for $\sigma_{AB}$ to correspond to a physical quantum state, namely to satisfy the uncertainty principle, the following condition has to be satisfied:
\begin{equation}\label{gaussstate}
{ \sigma}_{AB} + i\Omega^{(n)}\geqslant 0 \, ,
\end{equation}
where $n=n_A+n_B$ and the inequality means that the matrix in the l.h.s. is positive semi-definite.

\subsubsection{Two-mode entangled states}
In the following we consider two main classes of Gaussian states: the two-mode squeezed states  \cite{review1} and the three-mode GHZ/W states \cite{ADESSOJPA07}, 
the corresponding covariance matrices being indicated, respectively, as $\sigma_{2,r}$ and  $\sigma_{3,r}$. 
First, in order to define $\sigma_{2,r}$, consider a two-mode scenario, where Alice and Bob own each a single mode ($n_A=n_B=1$). 
The covariance matrix $\sigma_{2,r}$ corresponding to a two-mode squeezed state is given by Eq. (\ref{sigmaAB}), 
with $  A=  B=\cosh(2r) \,  {I} $ and $  C=\sinh(2r) \,   Z$, where $ {I}$ is the identity matrix and $  Z=\mbox{diag}(1,-1)$, namely:
\begin{eqnarray}\label{sigma0-2modes}
\sigma_{2,r}  = 
\left( \begin{array}{ccc}
\cosh(2r) \, {I} & \sinh(2r)  \, {Z} \\
 \sinh(2r)  \, {Z}  & \cosh(2r)  \, {I}
\end{array} 
\right) \, ,
\end{eqnarray}
 As the squeezing parameter $r\geqslant 0$ increases, the two modes become more and more correlated (entangled) \cite{review1}, where $r=0$ corresponds to a separable state, namely the two-mode vacuum state. The maximally entangled EPR state corresponds to $\lim_{r\rightarrow \infty} \sigma_{2,r}$. It must be noticed that $\sigma_{2,\infty}$ corresponds to an infinite energy state \cite{review1} and therefore it cannot be realized experimentally.

\subsubsection{Three-mode entangled states}

The GHZ/W state  is a three-mode Gaussian state which is entangled among each mode. It is realized by three squeezed beams mixed in a tritter \cite{ADESSOJPA07}. In case of equal squeezings, the corresponding covariance matrix is given by
\begin{eqnarray}\label{EPR}
\sigma_{3,r}  = 
\left( \begin{array}{ccc}
\sigma(r) & \epsilon(r) & \boldsymbol{}\epsilon(r) \\
\epsilon(r)  & \sigma(r)& \epsilon(r) \\
 \epsilon(r) & \epsilon(r) &\sigma(r)
\end{array} 
\right) \, ,
\end{eqnarray}
where 
$
\sigma(r) = \mbox{diag}((\expp{2r}+2\expp{-2r})/3,(\expp{-2r}+2\expp{2r})/3 ) $ and 
$ \epsilon(r) = (2/3) \sinh(2r)   {Z} $. The parameter $r\geqslant 0$ is the global squeezing parameter of the state, where $r=0$ corresponds to the separable case and   $r\rightarrow \infty$ provides maximal entanglement.

Consider the bipartite scenario where Alice owns the first two modes of $\sigma_{3,r}$, while Bob owns the last mode. The covariance matrix $\sigma_{3,r}$ can be divided into blocks as in Eq. (\ref{sigmaAB}), where 
$$  A=\left( \begin{array}{ccc}
\sigma(r) & \epsilon(r)\\
\epsilon(r)  & \sigma(r) 
\end{array} 
\right)\,,\,\,\,   B =\sigma(r) \,,\,\,\,   C = \left( \begin{array}{ccc}
\epsilon(r)\\
\epsilon(r) 
\end{array} 
\right) \, .
$$

\section{Gaussian channels and evolutions}\label{gevo}
Gaussian channels are those that preserve Gaussianity of quantum states, that is, they map Gaussian states into Gaussian states. They can be fully characterised by their application on the displacement vector and the covariance matrix. 
Nonetheless,
since we are interested only in the information contained in the covariance matrix of Gaussian states, we represent the action of a generic Gaussian transformation as \cite{review1}
\begin{align}
\label{transf}
    \Lambda: \sigma \, \mapsto \, \sigma' &=  T\sigma   T^T +   N \, ,
\end{align}
where $T$ and $N$ are $2n\times 2n$ matrices of reals. 
Moreover, $N$ must be symmetric to preserve the symmetry of the covariance matrices. Thus, any Gaussian channel can be represented by the pair $\Lambda\equiv(T,N)$, such that $\sigma^{\prime} = (N,T)\sigma$.
It is clear from Eq. (\ref{transf}) that the identity map corresponds to $(  T,   N) =(   I,   {0})$, where, again, $I$ is the identity matrix and ${0}$ is the null matrix.
 
The channel 
$(T,N)$
CPTP if and only if 
\cite{Lindblad_2000}
\begin{equation}\label{CPTP}
  N  - i   T\Omega   T^T + i\Omega\geqslant 0 \, .
\end{equation}
For single-mode channels, condition (\ref{CPTP}) reduces to the following two conditions 
\begin{eqnarray}\label{N1mode}
N&=& N^T \geqslant 0 \, \\
\det N &\geq& (\det T -1 )^2  \label{CPTP1mode}\, .
\end{eqnarray}

A Gaussian {dynamical evolution} can be denoted by the time-parametrised family $\{\Lambda_t\}_{t\geq 0}=\{T_t,N_t\}_{t\geq 0}$, where the channel $(T_t,N_t)$ that represents the evolution at time $t$ is called dynamical map. One expects that, naturally, at $t=0$ the dynamics is given by the identity channel, i.e.,  $(  T_0,   N_0)= (  I,    {0})$.
We further assume the evolution to be divisible, i.e.,  for any arbitrary times $0\leqslant s\leqslant t$ one can write
\begin{equation}\label{compevo}
(  T_t,  N_t)= (  T_{t,s},  N_{t,s}) \circ  (  T_s,  N_s)= (  T_{t,s}   T_s,   T_{t,s}   N_s   T_{t,s}^T+  N_{t,s}) \, ,
\end{equation}
where we used the composition law for Gaussian channels \footnote{This composition law can be obtained by applying Eq. (\ref{transf}) for  $\sigma \mapsto \sigma'=(T',N')\sigma $ and again for $\sigma'\mapsto \sigma''=(T'',N'')\sigma'$. }
$(  T'',  N'') \circ  (  T',  N')= (  T''   T',   T''  N'  T''^T+  N'') $ and
$(T_{t,s},N_{t,s})$ is called the intermediate map of the evolution for the time interval $[s,t]$.
Notice that $  N_{t,s}$ has to be symmetric. 
Importantly, for a general evolution the intermediate map $(  T_{t,s},  N_{t,s})$ could  be non-CPTP for some $0<s\leqslant t$. 
This fact can be used to define \textit{Markovian} Gaussian evolutions as the CP-divisible family of Gaussian channels $\{  T_t,  N_t\}_{t \geqslant 0}$. In other words, Markovian  evolutions   are those  with   CPTP intermediate maps $(  T_{t,s},  N_{t,s})$ for all $0< s\leqslant t$. In case the evolution is not CP-divisible, we call it \textit{non-Markovian}.

In the following we consider Gaussian evolutions that are applied only to one mode of a multimode Gaussian system. Accordingly, the covariance matrix of such multimode Gaussian state evolves as
\begin{eqnarray}
\label{eq:local-evolution}
\sigma(t) &=&(   T_t^{(1)} \oplus   I,    N_t^{(1)} \oplus    0) \sigma(0) \nonumber \\ 
&=&(   T_t^{(1)} \oplus   I )\sigma(0)   (   T_t^{(1)} \oplus   I )^T +   N_t^{(1)} \oplus    0 \label{transft} \, ,
\end{eqnarray}
with $(T_t^{(1)},N_t^{(1)})$ being a single-mode dynamical Gaussian channel. In the remainder of this text, we drop the label $^{(1)}$ to lighten our notation. 
We follow by describing how information quantifiers can be used to witness non-Markovianity of evolutions through their non-monotonic behaviors, namely backflows.

\section{Non-Markovianity witnesses}
\label{sec:NMwitnesses}

Given a functional $\mathcal{I}:S(\mathcal{H} )\rightarrow \mathbb{R}^+$ that maps quantum states into non-negative real numbers, we call it an \textit{information quantifier} if it is non-increasing under CPTP maps, namely if $\mathcal{I}(\hat\rho)\geqslant \mathcal{I}(\Lambda(\hat\rho)) $ for all CPTP maps $\Lambda:S(\mathcal{H} )\rightarrow S(\mathcal{H} )$ and states $\hat\rho\in S(\mathcal{H} )$.
The minimum value $\mathcal{I}=0$ is interpreted as the absence of the considered information in the state.
It follows that all evolutions $\{\Lambda_t\}_{t\geq 0}$ cannot increase the amount of information contained in the initial state $\hat{\rho}(0)$, namely $\mathcal{I}(\hat\rho(0))\geqslant \mathcal{I}(\hat\rho(t))$ for all $t \geqslant 0$ and $\hat\rho(0)\in S(\mathcal{H})$, where $\hat\rho(t):=\Lambda_t(\hat\rho(0))$.
 Nonetheless, it could be the case that an intermediate map between two times $s< t$ is not CPTP and that we obtain the increase $\mathcal{I}(\hat\rho(s))< \mathcal{I}(\hat\rho(t))$. Hence, since Markovian evolutions are characterized by having CPTP intermediate maps, any increase, or \textit{backflow}, of $\mathcal{I}$ \textit{witnesses} non-Markovianity.

Notice that, in general, it does not suffice to consider the evolution of a single initial state $\hat\rho(0)$ in order to state that an evolution is Markovian due to the monotonicity of $\mathcal{I}(\hat\rho(t))$. Indeed, even if a state $\hat\rho(0)$ does not allow observing backflows of $\mathcal{I}$, there may be a different state $\hat\rho'(0)$ for which an increase of $\mathcal{I}(\hat\rho'(t))$ can be observed. The same is true for $\mathcal{I}$: some quantifiers are not able to witness certain types of non-Markovian evolutions.

In this context, a key ingredient is the use of ancillary systems. Indeed, we can use initial states $\hat    \rho(0) \in S(\mathcal H \otimes \mathcal H')$ such that $\hat \rho(t) = \Lambda_t\otimes I' (\hat \rho(0))$, where the non-evolving ancillary system is defined over the Hilbert space $\mathcal H'$ and $I'$ is the identity map on $S(\mathcal H')$.
In general, initializations that make use of ancillas allow witnessing non-Markovianity with higher precision. Indeed, for some evolutions and information quantifiers, we can obtain backflows if and only if particular system-ancilla initializations are considered  \cite{BognaPRL,witnessbackflow,desantis2020correlations}. Moreover, the dimension of the ancilla is also important: depending on the case, a minimal ancillary size could be required to obtain backflows.

In the following, we exploit two information quantifiers as non-Markovianity witnesses: Gaussian steerability and entanglement. Our results reveal that there exist some non-Markovian Gaussian evolutions that   (i) cannot be witnessed by means of the aforementioned correlations with \textit{any} two-mode Gaussian initialization, but  (ii) can be witnessed by using three-mode initial Gaussian states, where in (i) and (ii) we respectively consider one and two ancillary modes. Thus, we highlight the crucial role that ancillary modes can play.

More in details, we consider the scenario where Alice and Bob share a Gaussian correlated system $A_1A_2B$, where $A_1$ is Alice's evolving system, $A_2$ is Alice's ancillary system (in case there is one) and $B$ is Bob's ancillary system. Hence, we compare the potentials of the settings $A_1|B$ and $A_1A_2|B$ to provide correlation backflows, where $A_1$, $A_2$ and $B$ are one-mode systems. As described before, the aim of this work is to describe the advantages of using the three-mode setup $A_1A_2|B$.

Finally, we propose the following analogy between finite and infinite dimensional evolutions to discuss the minimal ancillary sizes of $A_2$ and $B$ needed to observe correlation backflows when $A_1$ is a generic $n$-mode Gaussian evolving system. There exists a hierarchy for the degree of non-Markovianity of $d$-dimensional evolutions called $k$-divisibility \cite{degreeofNM}, which is based on the minimal ancillary dimension needed to obtain information backflows \cite{witnessbackflow}.  The backflows considered here correspond to increases in  distinguishability of two states  given with a-priori probabilities $p$ and $1-p$, which are defined over the evolving system and an ancilla. If an invertible evolution is $k$-divisible but not $k+1$-divisible, we can obtain backflows if and only if $k+1$-dimensional (or larger) ancillas are considered. In this framework, Markovianity corresponds to $d$-divisibility. Hence, $d$ is the largest ancillary dimension needed to witness non-Markovianity, which is required for  $d-1$-divisible evolutions. 
In terms of correlation backflows, by considering the setting $A_1A_2|B$ explained above, $d+1$ and $2$ are, respectively, the minimum dimensions of $A_2$ and $B$ that have been proven to be sufficient to witness any invertible non-Markovian evolution \cite{DDSshort}.

Similarly, in order to observe correlation backflows from $n$-mode non-Markovian Gaussian evolutions, we may expect to need $n+1$-mode $A_2$ ancillas. Nonetheless, Gaussian non-Markovianity  follows a simpler  hierarchy \cite{PhysRevLett.118.050401}: intermediate maps are either CP, positive or non-positive, where Markovianity corresponds to 1-divisibility. Therefore, we expect that, given a generic invertible non-Markovian Gaussian evolution, a minimal requirement for a bipartite system $A_1A_2|B$ to provide correlation backflows is that $A_2$ and $B$ are respectively (at most) two-mode and one-mode Gaussian systems, no matter the number of evolving modes of $A_1$.

\subsection{Gaussian steerability}
Gaussian steerability is a form of quantum correlations and, similarly to other correlation measures, is non-increasing under the action of CPTP local maps. 
For instance, it means that if one is interested in a single mode dynamical channel $(T_t,N_t)$, one can construct a local evolution as in \eqref{eq:local-evolution} 
and Gaussian steerability can be considered as an information quantifier and used to witness non-Markovianity through backflows, as suggested in Ref. \cite{ParisCV}. 
There, the authors show that $\sigma_{AB}(0)=\sigma_{2,r}$ can be deployed as initial state to witness  non-Markovianity for the quantum Brownian motion model---see the description in Section \ref{BrownMot}. 
In what follows, we first present the mathematical description of Gaussian steerability. 
Then, we provide examples of dynamics where this correlation cannot witness non-Markovianity when using \textit{any} two-mode Gaussian system. 
We then proceed by showing that using three-modes one can witness non-Markovianity in many cases where two modes fail. 
We furthermore show that even with three entangled modes it can happen that  some non-Markovian evolutions cannot be witnessed using Gaussian steerability.

Consider a bipartite scenario where Alice and Bob share an $n_A+n_B$-mode Gaussian state with covariance matrix $\sigma_{AB}$, where Alice holds the first $n_A$ modes and Bob holds the last $n_B$ modes---see Eq.~\eqref{sigmaAB}.
A quantifier for the potential of Alice to steer Bob's share through Gaussian measurements has been introduced in Ref. \cite{ADESSOgsteering}. 
It turns out that $\sigma_{AB}$ is Gaussian steerable from Alice to Bob, or $A\rightarrow B$ steerable with Gaussian measurements, if and only if the following condition is violated \cite{WISEMAN1}:
\begin{equation}\label{ineqwiseman}
\sigma_{AB} + i (  0_A \oplus \Omega_B )\geqslant 0 \, ,
\end{equation}
where $  0_A$ is the $2n_A\times 2n_A$ null matrix and $\Omega_B=\Omega^{(n_B)}$. This condition is equivalent to the Schur complement of $B$
\begin{equation}\label{Schur}
 {M}^\sigma_B =   B -   C^T   A^{-1}   C \, ,
\end{equation}
 not being a physical covariance matrix; that is a violation of the following inequality
\begin{equation}\label{Schurphys}
 {M}^\sigma_B + i \Omega_B \geqslant 0 \, .
\end{equation}
Therefore, ${A\rightarrow B}$ Gaussian steerability can be verified by studying the set $\{\nu_i\}_{i=1}^{n_B}$ of \textit{symplectic eigenvalues}  of $  M^\sigma_B$. 
Recall that these are associated to the absolute value of the eigenvalues $ \{\pm \nu_i\}_{i=1}^{n_B}$ of the matrix $i\Omega_B   M_B^{\sigma}$ \cite{Serafini}. 
It can be shown that Eq. (\ref{Schurphys}) is violated if and only if $\nu_i< 1$ for one or more $i$  \cite{review1}. Hence, following Ref. \cite{ADESSOgsteering}, one can  quantify $A\rightarrow B$ Gaussian steerability as:
\begin{equation}\label{gsteering}
\mathcal{G}_{A\rightarrow B}(\sigma_{AB}) = \max\left\{0 , - \sum_{\nu_i<1} \log \nu_i \right\} \, .
\end{equation}
In case we want to evaluate ${ B\rightarrow A}$ Gaussian steerability, we replace $ {M}^\sigma_B$ with the Schur complement of $A$, namely $ {M}^\sigma_A =   A -   C   B^{-1}   C^T $, and evaluate its symplectic eigenvalues. Notice that in general steering is not symmetric, i.e.,  $\mathcal{G}_{A\rightarrow B}(\sigma_{AB})\neq \mathcal{G}_{B\rightarrow A}(\sigma_{AB})$.

\subsubsection*{Measurement incompatibility and steering}
A necessary condition for Gaussian steerability is given by Gaussian measurement incompatibility. 
Imagine an $A\rightarrow B$ Gaussian steering scenario, where Alice owns $n_A$ Gaussian modes which are transformed by the Gaussian channel $(T,N)$. 
In case the action of the (dual) channel $(  T,   N)^*$ makes the set of Alice's Gaussian measurements compatible, no $A\rightarrow B$ Gaussian steering can be performed ---that is ${\cal G}_{A\to B}((T\oplus I_B,N\oplus 0_B)\sigma_{AB})=0$ for all initializations $\sigma_{AB}$. 
In turn, a Gaussian channel breaks incompatibility of all Gaussian measurements if and only if \cite{TEIKOINCOMP,PhysRevA.96.042331,INCOMPBREAK}
\begin{equation}\label{incompbreak}
  N - i   T \Omega   T^T \geqslant 0 \, . 
\end{equation}
We call a channel $(  T,   N)$ \textit{ Gaussian incompatibility breaking} (GIB) in case it satisfies Eq. (\ref{incompbreak}).
This immediately leads to the following observation:

{\it Observation 1.---}Consider a dynamics that breaks incompatibility of all Gaussian measurements on Alice's side within some time interval $t\in(t_{1},t_{2})$.  Any non-Markovian behaviour, namely the violation of the CP-divisibility condition, of the dynamics within this interval cannot be witnessed by Gaussian steerability from Alice to Bob. Indeed, steering is equal to zero in the time interval $(t_{1},t_{2})$. 

On the other hand, Alice can always extend her system to include one or more new Gaussian modes \textit{which do not undergo the Gaussian channel}. %That is, her total share undergoes the channel $(T_t\oplus{I},N_t\oplus{0})$.
Such an extension leads to our second observation:

{\it Observation 2.---}If one or more of Alice's modes do not undergo the Gaussian dynamics, namely if Alice extends her modes by at least one such that her total share undergoes the Gaussian dynamics $(T_t\oplus{I},N_t\oplus{0})$, the criterion \eqref{incompbreak} is always violated, i.e.,  the dynamics on Alice is never GIB.

Note that Observation {2} does not imply that Gaussian steerability from Alice's extended system to Bob can witness non-Markovianity. 
Indeed, on the one hand measurement incompatibility is a necessary but not sufficient condition for Gaussian steerability. 
On the other hand, provided that one has non-zero steering, it is not guaranteed that steering backflows  are always observed when non-Markovianity is at play.
Nonetheless, we can increase the chance to witness a bigger class of non-Markovian dynamics by simply extending the number of modes. We showcase this through some examples in Section Sections \ref{GaussSteerClNoise} and \ref{GaussSteerLossy}. 
%
%
% %
\subsection{Entanglement}
A necessary condition for the separability of a bipartite $n_A+n_B$-mode Gaussian state $\hat\rho_{AB}$ with covariance matrix $\sigma_{AB}$, is given by the positivity of the partial transposition of the density matrix, namely the PPT condition \cite{PeresPPT,PPT2}, which states that separable states satisfy the condition:
\begin{equation}\label{sep}
\sigma_{AB}+i \Omega_A \oplus \Omega_B^T \geqslant 0 \, ,
\end{equation}
where $\Omega_A=\Omega^{(n_A)}$ and $\Omega_B=\Omega^{(n_B)}$. Hence, a quantifier for the entanglement in $\sigma_{AB}$ can be defined as:
\begin{equation}\label{gent}
\mathcal{E}_{PPT}(\sigma_{AB}) = \max\left\{0 , - \sum_{\mu_i<0} \mu_i \right\} \, , 
\end{equation}
where $\mu_i$ is the $i$-th eigenvalue of $\sigma_{AB}+i \Omega_A \oplus \Omega_B^T$.
The PPT condition \eqref{sep} is a necessary separability condition in general, but turns out to be sufficient for any $1+n_B$-mode  and $n_A+1$-mode Gaussian state, namely when at least one of the two parties is single mode  (the case we will be considering in Sec.~\ref{sec:examples}). 

A Gaussian channel $(T,N)$ applied to Alice's share is entanglement breaking (EB), i.e., nullifies the entanglement content of any bipartite input state, if and only if the matrix $N$ admits~\cite{holevo2008entanglement}
\begin{align}\label{eq:Gauss_EB}
    N=N_1+N_2,\hspace{.2cm}\text{where}\hspace{.2cm} 
    N_1\geqslant i\Omega^{(n_A)},\hspace{.3cm} N_2\geqslant iT \Omega^{(n_A)} T^T.
\end{align}
{\it Remark.---}Any 
EB channel is also GIB. 
To see this for Gaussian channels considered here, note that \eqref{eq:Gauss_EB} implies that a necessary condition for EB is to have $N_1\geqslant 0$. When we add this to the condition for $N_2$, we revive \eqref{incompbreak}. Also, notice that the reverse is not necessarily true.

We can now make the following two observations analogous to Observations 1 and 2.

{\it Observation 3.---}Consider a dynamics on Alice's side that is EB within some time interval $t\in(t_{1},t_{2})$. Any non-Markovian behaviour, namely the violation of the CP-divisibility condition, of the dynamics within this interval cannot be witnessed by entanglement between Alice's and Bob's system. Indeed, entanglement is always zero in  $(t_1,t_2)$.

{\it Observation 4.---}If one or more of Alice's modes do not undergo the Gaussian dynamics---i.e.,  if Alice extends her modes by at least one, such that her total share undergoes the Gaussian channel $(T_t\oplus{I},N_t\oplus{0})$---the criterion \eqref{eq:Gauss_EB} is always violated, i.e.,  the dynamics on Alice is never EB.

It turns out that
for a generic one-mode Gaussian channel $(  T,   N)$, 
the EB character can be tested by applying the channel locally over  
the maximally entangled two-mode squeezed state $\sigma_{2,\infty}=\lim_{r\rightarrow \infty} \sigma_{2,r}$, i.e.,  the separability condition $\mathcal{E}_{PPT}((  T\oplus   I,   N \oplus    0) \sigma_{2,\infty})=0$ is a necessary and sufficient condition for the EB character of the channel \cite{Serafini}.

\section{Paradigmatic examples}
\label{sec:examples}
\begin{figure}
\centering
\begin{overpic}[width=.4\textwidth]{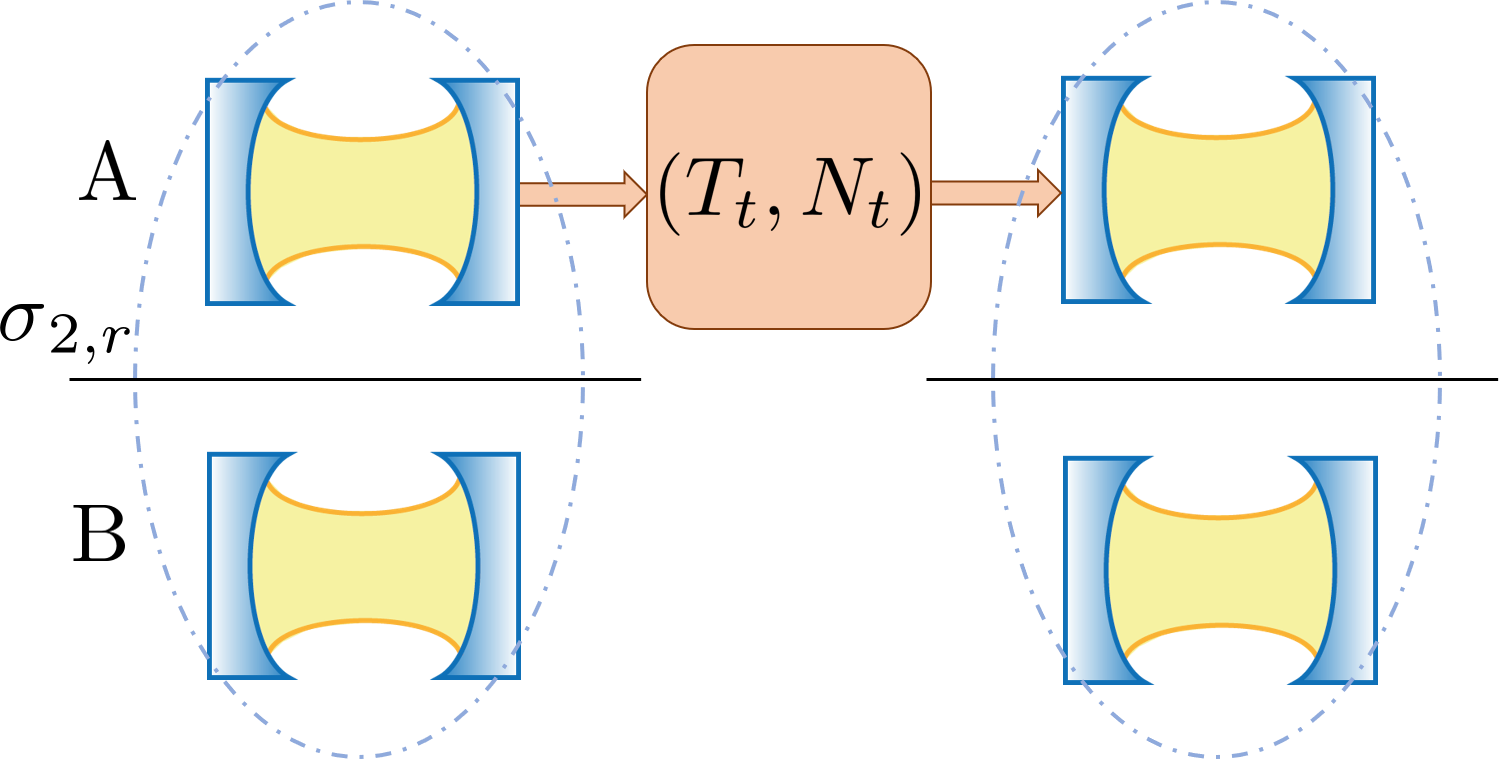}
\put(1,50){\bf (a)}
\end{overpic}
\vspace{.3cm}
\\
\begin{overpic}[width=.4\textwidth]{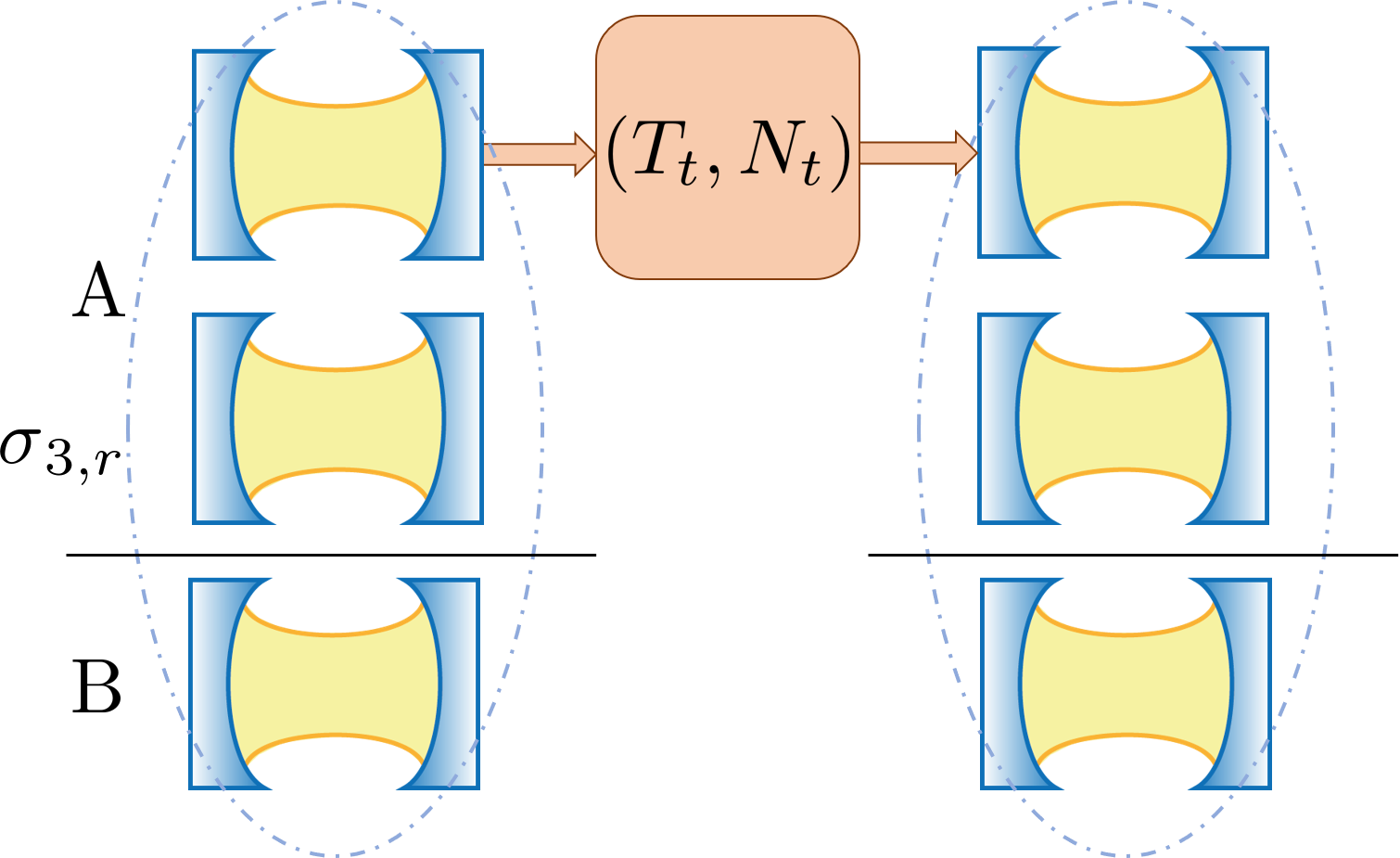}
\put(1,60){\bf (b)}
\end{overpic}
\\
\caption{
{
\label{scheme} 
Schematic setup for the detection of non-Markovianity
with (a)
two-mode squeezed input state 
and
with (b)
GHZ/W three-mode input state.
Notice the selected bipartition concerning Alice's and Bob's shares (black horizontal line) where only Alice is allowed to own more than one mode.
We consider Gaussian evolutions $(T_t, N_t)$ that are applied only to the first mode of Alice's share.
Non-monotonic behaviors of Gaussian steering/entanglement quantifiers as function of time (backflows) are used as non-Markovianity witnesses. 
The GHZ/W three-mode configuration (b) can activate the potential to witness non-Markovian behaviors which in the two-mode scenario (a) do not imply any revivals of quantum correlations.  
}
}
\end{figure}

Here, we demonstrate how by using an extra auxiliary mode on Alice's share, one can witness non-Markovianity through correlation backflows within a bigger class of Gaussian dynamics, if compared to the one-ancillary mode scenario.
The potential of the method was anticipated in Observations 2 and 4: since some of the modes owned by Alice do not undergo the channel and can maintain their correlations with Bob's side the correlation quantifier does not nullify implying more chances of observing its backflow.
 A schematic representation of our settings is reported in Fig.~\ref{scheme}.
\subsection{Classical noise channel}
\label{sec:classical-noise}
We start by probably the simplest example, i.e.,  a classical noise channel applied to a single mode.
It is given by \cite{review1} 
\begin{equation}
\label{toy-channel}
(T_t,N_t)=(I,\eta(t) I)~,
\end{equation}
where $\eta(t)\geqslant 0$ is a time-dependent continuous function such that $\eta(0)=0$.
The form for the intermediate map of this evolution can be derived using Eq. (\ref{compevo}), obtaining, for any $s,t$ such that $0<s<t$,
\begin{equation}\label{noiseint}
(  T_{t,s} ,  N_{t,s})= (   I, (\eta(t)-\eta(s))   I) \, .
\end{equation}
It follows that this Gaussian evolution is Markovian if and only if $\eta(t)$ is monotonically increasing. Indeed, from Eq. (\ref{N1mode}), 
as soon as $\eta(t)-\eta(s)<0$ the corresponding intermediate channel is non-CPTP.
\subsubsection{Gaussian steerability}\label{GaussSteerClNoise}
Consider a two-mode Gaussian state shared between Alice and Bob described by the covariance matrix ${\sigma_{AB}}$. When acting on Alice's mode, the dynamical map $(I,\eta(t)I)$ makes the set of all Gaussian measurements compatible if and only if $\eta(t)\geqslant 1$ (see Eq.~\eqref{incompbreak} or \cite{INCOMPBREAK}). Therefore, according to Observation 1, it is not possible to obtain information backflows through Gaussian steerability $\mathcal{G}_{A\rightarrow B}$ at times $\{\,t~|\,\eta(t)\geqslant 1\}$. However, inspired by Observation 2, Alice can extend her system to contain an auxiliary mode, 
such that her share undergoes a local evolution on the first mode, see Eq.~\eqref{eq:local-evolution}. For instance, if we take $\sigma_{AB}={\sigma_{3,r}}$, where Alice holds the first two modes, there will be a backflow in ${\cal G}_{A\to B}$ at anytime that $\eta(t)$ decreases, i.e.,  whenever the dynamics is not CP-divisible. Interestingly, this is true for any squeezing $r>0$. We provide the proof of this result in Appendix \ref{appendix:classical}, where we obtain the analytical forms of $\mathcal{G}_{A\rightarrow B}(\sigma_{2,r}(t))$ and $\mathcal{G}_{A\rightarrow B}(\sigma_{3,r}(t))$ for any $r$ and $\eta(t)$.

As an example, let the noise assume the following time dependence 
%
%in this regime. By considering  $\sigma_{2,2}$ as initial state for different shapes of non-Markovian $n(t)$, we study
\begin{equation}\label{n(t)decaying}
    \eta(t)=t^2/(t^2-2t+2)
    .
\end{equation} 
Since $\eta(t)$ is monotonically decreasing for $t\geq 2$, it follows that the evolution is non-Markovian and the intermediate map $(  T_{t,s},   N_{t,s})$ is not CPTP for any $2\leqslant s<t$. Nonetheless,  $( T_{t},   N_{t})$ breaks the incompatibility of all Gaussian measurements for $t\geqslant 1$ and therefore no backflow of $\mathcal{G}_{A\rightarrow B}$ can be observed when Alice holds only one mode. This is showcased in Fig. \ref{FIGtoy}(a) where we take $\sigma_{AB}=\sigma_{2,r}$. This is contrary to when we take $\sigma_{AB}={\sigma_{3,r}}$, where Alice holds the first two modes. As seen from Fig. \ref{FIGtoy}(a), in this case $\mathcal{G}_{A\rightarrow B}$ increases whenever the dynamics is not CP-divisible.

\subsubsection{Entanglement}
\begin{figure*}
\centering
\begin{overpic}[width=.45\linewidth]{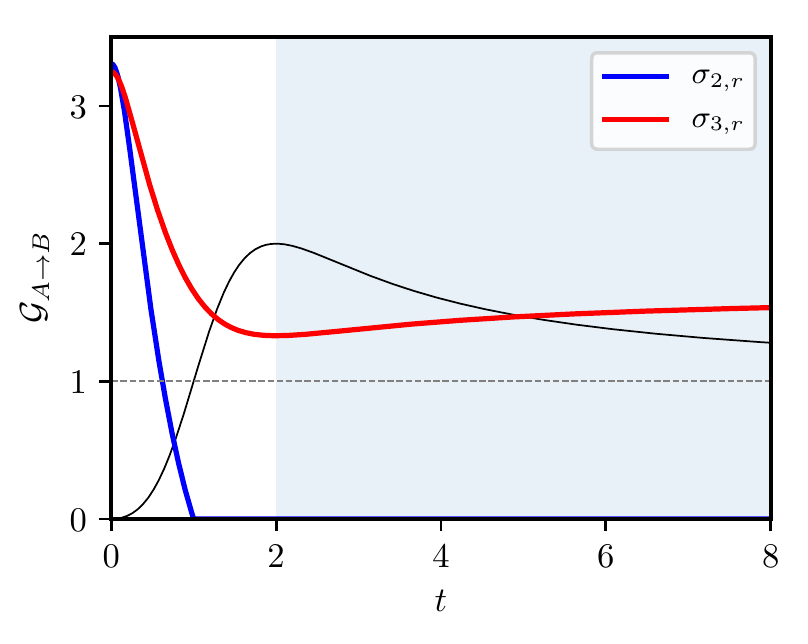}
\put(20,65){\bf (a)}
\end{overpic}
\qquad
\begin{overpic}[width=.45\linewidth]{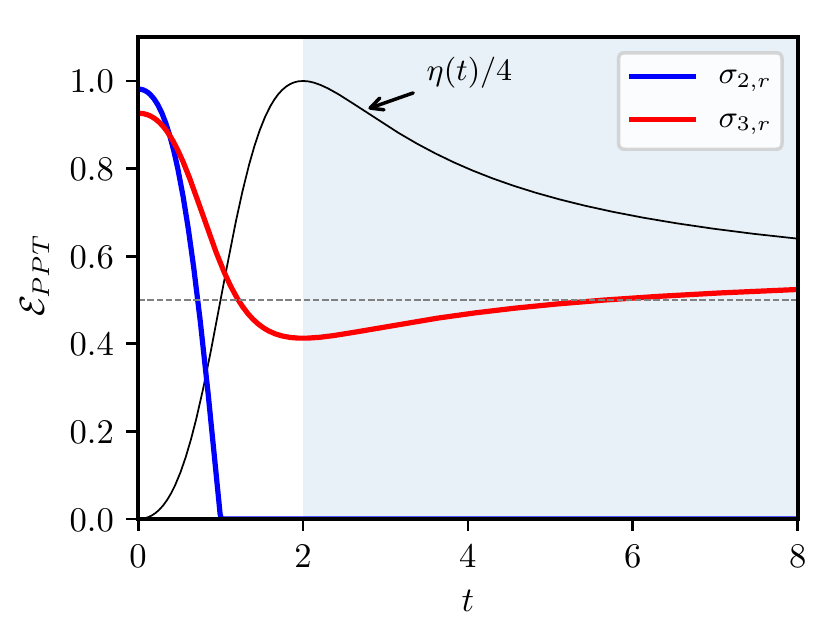}
\put(23,65){\bf (b)}
\end{overpic}
\\
\caption{\label{FIGtoy} 
(a)
Gaussian steerability $\mathcal{G}_{A\rightarrow B}$ as function of time for
the three-mode squeezed state initialization $\sigma_{3,r}(0)$  
and 
the two-mode squeezed state initialization $\sigma_{2,r}(0)$  
and Alice's first mode subjected to the classical noise evolution with
$\eta(t)=t^2/(t^2-2t+2)$ (black line, Eq.~\eqref{n(t)decaying}), 
i.e.
such that $\eta(0)=0$, $\eta(1)=1$ (horizontal dashed line, level of noise for which we have the GIB property), $\eta(2)=2$, $\eta(\infty)=1$, it is increasing in $[0,2]$ and decreasing in $[2,\infty]$ (non-Markovian interval, blue shadow region).
We cannot observe any backflow of $\mathcal{G}_{A\rightarrow B}(\sigma_{2,r}(t))$ (blue line) in the time interval when the evolution is non-Markovian, because the dynamical maps $(T_t,N_t)$ of the evolution are GIB for $t\geq1$. 
We do observe, instead, a backflow of $\mathcal{G}_{A\rightarrow B}(\sigma_{3,r}(t))$ (red line) in the time interval when the evolution is non-Markovian. 
The plots are made for $r=2$. 
(b)
Same as in (a) but for $\mathcal{E}_{PPT}$ and for
$\eta(t)=2 t^2/(t^2-2t+2)$ (Eq.~\eqref{func-toy-ent}), such that $\eta(0)=0$, $\eta(1)=2$ (horizontal dashed line, level of noise for which we have the EB property), $\eta(2)=4$, $\eta(\infty)=2$, it is increasing in $[0,2]$ and decreasing in $[2,\infty]$.
For convenience we report now the rescaled quantity $\eta(t)/4$, black line.
While we cannot observe any backflow of $\mathcal{E}_{PPT}(\sigma_{2,r}(t))$ (blue line) in the time interval when the evolution is non-Markovian, namely, again, for $t\geqslant 2$ (because the dynamical maps $(T_t,N_t)$ of the evolution are EB for $t\geq1$), 
we do observe, instead, a backflow of $\mathcal{E}_{PPT}(\sigma_{3,r}(t))$ (red line).
}
\end{figure*}

The dynamical map $(I,\eta(t)I)$ breaks the entanglement of all states if and only if $\eta(t)\geqslant 2$, see Eq.~\eqref{eq:Gauss_EB}.
Accordingly,  
we can choose $\eta(t)$ to be the following time-dependent function,
\begin{equation}\label{func-toy-ent}
\eta(t)=2 t^2/(t^2-2t+2)~,    
\end{equation}
namely the function \eqref{n(t)decaying} rescaled by a factor $2$, implying that the non-Markovian interval is the same as before, i.e., $[2,\infty]$ where the function is decreasing.
For $t\geq t_{EB}=1$ we have the EB property ($\eta(1)=2$ and $\eta(t)\geq 2$ if and only if $t\geq1$).
Analogously to what we observed for steering,
in Fig.~\ref{FIGtoy}(b) we show that
whether we cannot observe any backflow of $\mathcal{E}_{PPT}(\sigma_{2,r}(t))$ (this quantity nullifies for $t\geq t_{EB}$), 
we do observe, instead, a backflow of $\mathcal{E}_{PPT}(\sigma_{3,r}(t))$ as soon as the evolution becomes non-Markovian.
\subsection{Lossy channel and the quantum Brownian motion}\label{BrownMot}
The second example we consider is  the wider  class of lossy channels, which are the evolutions with dynamical maps
\begin{equation}
\label{lossy-channel}
(T_t,N_t)=(\tau(t)I,\eta(t) I) \, ,   
\end{equation}
including the classical noise channel \eqref{toy-channel} in the particular cases $\tau(t)=1$.
The intermediate maps of these evolutions assume the form
\begin{equation}\label{BMint}
(  T_{t,s} ,  N_{t,s})= ( \tau(t,s)   I,\eta(t,s)   I) \, ,
\end{equation}
where $\tau(t,s)=\tau(t)/\tau(s)$ and $\eta(t,s)=\eta(t) - \eta(s)( {\tau(t)}/{\tau(s)})^2 $.
As we show in Appendix \ref{appendix:lossy}, the lossy channel is not CP-divisible, namely the intermediate map $(T_{t+\epsilon,t},N_{t+\epsilon,t})$ is not CPTP, at times $t$ if and only if either one or both of the following inequalities are violated
\begin{align}\label{eq:NM_LOSSY}
    {\dot \eta}(t) - 2(\eta(t)\pm 1)\frac{\dot \tau(t)}{\tau(t)}\geqslant 0.
\end{align}

The class of lossy channels is relevant in the description of quantum Brownian motion, where
a harmonic oscillator with frequency $\omega_0$ undergoes a dissipative dynamics by interacting with a bosonic bath at temperature $T$. The total Hamiltonian is quadratic, which guarantees that the system undergoes a Gaussian dynamics---details of the interaction and the derivation of the dynamics are given in the Appendix \ref{appendix:meQBM}. 
The two dynamical parameters $\eta(t)$ and $\tau(t)$ are connected to the physical parameters describing the system and the bath as follows
\begin{eqnarray}
\label{eq:Delta-and-gamma}
\tau(t)&=&\exppp{-\int_0^t \mbox{d}s~ \gamma(s)/2 } \, , 
\nonumber \\ 
\eta(t)&=&\tau(t)^2\int_0^t  \mbox{d}s ~\Delta(s)/\tau(s)^2 \, , 
\nonumber \\
\gamma(t)&=&\alpha^2 \int_0^t \mbox{d}\tau \int_0^\infty \mbox{d}\omega J(\omega) \sin(\omega \tau) \sin(\omega_0 \tau) \, , 
 \\
\Delta(t)&=&\alpha^2 \int_0^t \mbox{d}\tau \int_0^\infty \mbox{d}\omega J(\omega) \coth \left(\frac{\omega}{2T}\right) \cos(\omega \tau) \cos(\omega_0 \tau) \, . 
\nonumber
\end{eqnarray}
Here, $\Delta(t)$ is the so called diffusion coefficient and $\gamma(t)$ is the damping coefficient. The parameter $\alpha$ quantifies the system-bath interaction strength. 

For our simulations, we follow Ref.~\cite{ParisCV} and choose a spectral density  $J(\omega)$ of the bath with a Lorentz-Drude cutoff
\begin{equation}
\label{eq:specdens}
    J(\omega)=\frac{2\omega^s}\pi \frac{\omega_c^{3-s}}{\omega_c^2+\omega^2} \, ,
\end{equation}
where $\omega_c$ is the cutoff frequency.
The parameter $s$ defines the \textit{ohmicity} of the spectral density: $s<1$ corresponds to the sub-Ohmic regime, $s=1$ to the Ohmic regime, and $s>1$ the super-Ohmic regime.

This model is non-Markovian, i.e., the infinitesimal intermediate map $(  T_{t+\epsilon,t} ,  N_{t+\epsilon,t})$ is not CPTP, for those times $t$ such that one or both of the following inequalities are violated
\begin{equation}
\label{eq:QBM-NM}
    \Delta(t)\pm\gamma(t)\geqslant 0,
\end{equation}
a situation that is typically encountered when the cutoff frequency is smaller than the characteristic frequency of the oscillator, i.e.,  $\omega_c<\omega_0$ (see, e.g., \cite{ParisCV}). The criterion~\eqref{eq:QBM-NM} above can be also found directly by substituting ~\eqref{eq:Delta-and-gamma} in \eqref{eq:NM_LOSSY}.
\begin{figure*}
\centering
\begin{overpic}[width=.45\linewidth]{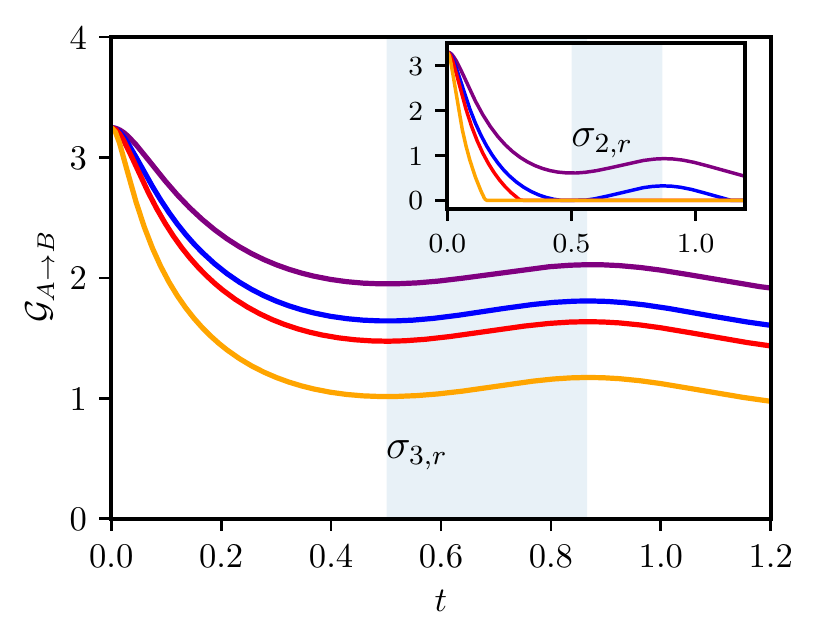}
\put(20,65){\bf (a)}
\end{overpic}
\qquad
\begin{overpic}[width=.45\linewidth]{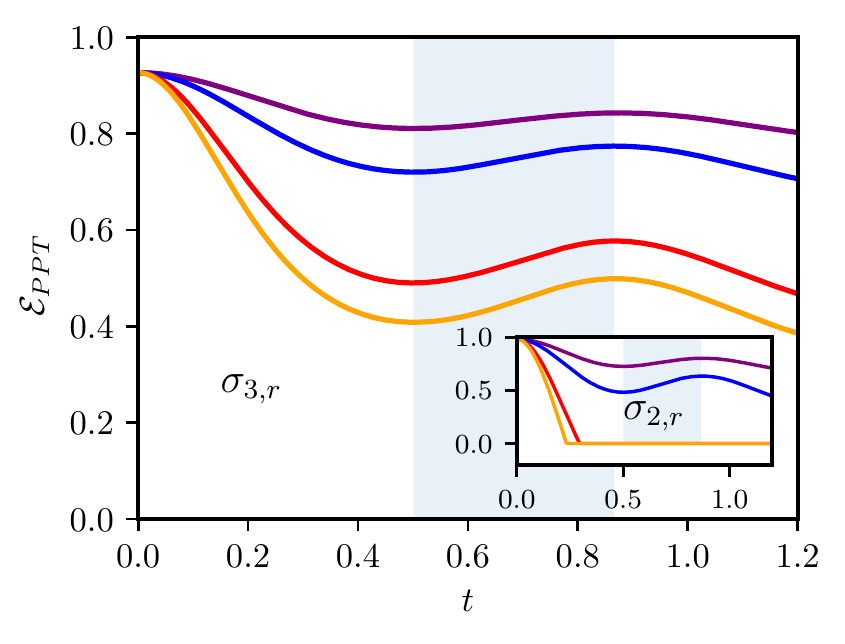}
\put(35,65){\bf (b)}
\end{overpic}
\\
\caption{
\label{FIGbrownian} 
(a)
Gaussian steerability $\mathcal{G}_{A\rightarrow B}$ as function of time for the three-mode squeezed state initialization $\sigma_{3,r}(0)$ and Alice's first mode subjected to the quantum Brownian motion,
for different values of the coupling parameter $\alpha$: 0.25 (purple), 0.35 (blue), 0.42 (red) and 0.7 (orange). 
Inset: $\mathcal{G}_{A\rightarrow B}$ for the two-mode squeezed state initialization $\sigma_{2,r}(0)$ with the same parameters. 
Here for $\alpha$ greater than $\approx 0.42$ (red), we cannot witness non-Markovianity as the dynamical maps of the evolution become GIB before any non-Markovian behavior. 
(b)
Same as in (a) but for 
entanglement $\mathcal{E}_{PPT}$ and for the following values of $\alpha$: 0.25 (purple); 0.35 (blue); 0.595(red), threshold value for
the $\mathcal{E}_{PPT}(\sigma_{2,r}(t))$ sensitivity; 0.7 (orange). 
Both $\mathcal{G}_{A\rightarrow B}(\sigma_{3,r}(t))$ and $\mathcal{E}_{PPT}(\sigma_{3,r}(t))$ show backflow whenever non-Markovianity is present (shadow blue regions, according to condition \eqref{eq:QBM-NM}). 
All the plots are made  in the Ohmic regime ($s=1$), setting $r=2$, $\omega_0=7$, $\omega_c=1$, $T=100$ (high temperatures).
}
\end{figure*}
%--------------------------------------------
%--------------------------------------------
\subsubsection{Gaussian steerability}\label{GaussSteerLossy}
Analogously to the classical noise channel, we first consider a two-mode state $\sigma_{AB}$, where Alice's share undergoes the dynamical map $(\tau(t)I,\eta(t)I)$. 
Ref.~\cite{ParisCV} uses such a setting to witness non-Markovianity of the channel in the context of quantum Brownian motion---where the initial state is $\sigma_{AB}=\sigma_{2,r}$. In several scenarios this technique allows to witness non-Markovianity via Gaussian steerability backflows.
However, under such a channel, all Gaussian measurements on Alice's mode become compatible if and only if $\eta(t)\geqslant \tau^2(t)$ (see Eq.~\eqref{incompbreak} or Ref.~\cite{INCOMPBREAK}).
Therefore, it is not possible to witness non-Markovianity through Gaussian steerability at times $\{t~|~\eta(t)\geqslant \tau(t)^2\}$. 
In Appendix \ref{appendix:lossy} we derive the analytical form of $\mathcal{G}_{A\rightarrow B}(\sigma_{2,r}(t))$ for any $r$, $\eta(t)$ and $\tau(t)$. Moreover, we show that a Gaussian steerability backflow is obtained, more precisely $\partial_t {\cal G}_{A\to B}>0$, if and only if the inequalities
\begin{align}\label{eq:GS_LOSSY_2m}
    {\dot \eta}(t)-2\eta(t)\frac{{\dot \tau}(t)}{\tau(t)} &< 0,\nonumber\\
    \eta(t) - \tau(t)^2 &<0,
\end{align}
are satisfied simultaneously (the latter inequality being the aforementioned non-GIB condition). 
Notice that the first inequality is the arithmetic average of the two possible violations of \eqref{eq:NM_LOSSY}, and hence
more restrictive than \eqref{eq:NM_LOSSY}. 
This implies that some non-Markovian evolution  cannot be witnessed. 

Nonetheless, when Alice's share is extended to include an auxiliary mode that does not undergo the channel, i.e.,  the dynamical map is $(\tau(t)I\oplus I, \eta(t)I\oplus 0)$---the set of Gaussian measurements will remain incompatible at all times. In particular, let Alice and Bob share the three-mode squeezed state $\sigma_{AB}=\sigma_{3,r}$, where Alice holds the first two modes. 
In Appendix \ref{appendix:lossy} we derive the analytical form of $\mathcal{G}_{A\rightarrow B}(\sigma_{3,r}(t))$ for any $r$, $\eta(t)$ and $\tau(t)$. Moreover, we show that we have a backflow $\partial_t {\cal G}_{A\to B}>0$ if and only if
\begin{align}\label{eq:GS_LOSSY_3m}
    {\dot \eta}(t)-2\eta(t)\frac{{\dot \tau}(t)}{\tau(t)} &< 0,
\end{align}
which for the quantum Brownian motion is equivalent to $\Delta(t) < 0$.  
This criterion is clearly less restrictive than \eqref{eq:GS_LOSSY_2m}---in that it does not require $\eta(t) < \tau^2(t)$. Thus, using the extra mode one can witness information backflow for a bigger class of dynamics. Notice that, the criterion~\eqref{eq:GS_LOSSY_3m}
is still more restrictive than ~\eqref{eq:NM_LOSSY}---except in case $\dot\tau(t)=0$, 
where $\tau(t)=1$ for all $t$ reduces the problem to the classical noise channel  (for quantum Brownian motion, instead, an analogous situation is realized in the limit of high temperatures in which $\vert\Delta(t)\vert\gg\vert\gamma(t)\vert~\forall t$, see discussion in Appendix \ref{appendix:low-temperature}).
In order to witness all non-Markovian Gaussian evolutions one might consider a different three-mode initialization or increase the number of ancillary modes. 
Yet, it is not guaranteed that by doing so one can witness all non-Markovian dynamics with Gaussian steerability.

In Fig.~\ref{FIGbrownian} (a), we plot Gaussian steerability ${\cal G}_{A\to B}$ for the quantum Brownian motion model at high temperatures. As a benchmark, we also depict the breakdown of CP-divisibility (shaded regions), i.e.,  the breakdown of inequality \eqref{eq:QBM-NM}.
We compare two cases with $\sigma_{AB}=\sigma_{2,r}$ and $\sigma_{AB}=\sigma_{3,r}$. 
As the inset shows, using the two-mode state is not successful for some of the parameter regimes that we consider. 
In particular, 
if  the interaction strength $\alpha$ is too large, the evolution of two-mode initializations cannot show any backflow of $A\rightarrow B$ Gaussian steerability because the evolution breaks the incompatibility of Alice's measurements. 
Furthermore, as one increases $\alpha$, the time intervals during which one cannot witness backflow of Gaussian steerability in the two-mode configuration increases.
On the contrary, in the three-mode scenario, if Alice owns the first two modes of $\sigma_{3,r}$, the incompatibility of Alice's measurements is not broken and we can witness non-Markovianity via backflows for all $\alpha$.

\subsubsection{Entanglement}\label{plzplz}
Firstly, notice that the dynamical map $(\tau(t)I,\eta(t)I)$ is EB iff $\eta(t)\geqslant \tau^2(t) + 1$ (see Eq.~\eqref{eq:Gauss_EB}).
When we use the two-mode squeezed state for small $r$, one can show that $\partial_t {\cal E}_{PPT}>0$ if the following two inequalities are satisfied simultaneously
\begin{align}
    \eta(t)-\tau^2(t)-1 & < 0 ,\\
    {\dot \eta}(t)-2(\eta(t)-1)\frac{{\dot \tau(t)}}{\tau(t)}&<0.
\end{align}
The first inequality expresses the fact that steerable states are entangled but in general not vice-versa,
i.e.,  entanglement is more {\it robust} to noise.
Interestingly,
provided steerability is non-zero,
the second inequality, being one of the two possible violations of \eqref{eq:NM_LOSSY},
when compared with \eqref{eq:GS_LOSSY_2m} suggests that, steerability (entanglement) could be preferable for observing backflows if $\dot{\tau}(t)/{\tau(t)}>$ $(<)$ $ 0$,
the two figures of merit yielding the same performance in the case $\dot{\tau}(t)=0$

While we are not able to retrieve analytical conclusions about entanglement backflows in the case of the three-mode initialization $\sigma_{3,r}$,
the potential of this configuration can be shown numerically for the quantum Brownian motion model at high temperatures.
In Fig.~\ref{FIGbrownian} (b), we plot ${\cal E}_{PPT}$ as function of time comparing the two cases  $\sigma_{2,r}(t)$ (inset) and $\sigma_{3,r}(t)$. 
Again, using the two-mode state is not successful  for large values of $\alpha$.
On the contrary, in the three-mode scenario, if Alice owns the first two modes of $\sigma_{3,r}$ we can witness non-Markovianity via backflows of entanglement for all $\alpha$.

\section{Discussion}
\label{sec:dicussion-conclusions}

We considered Gaussian steerability and entanglement as quantifiers of the correlations contained in a two-party Gaussian system and at the same time their backflows as non-Markovianity witnesses. 
We were interested in those non-Markovian evolutions that cannot be witnessed by backflows of these correlations when two-mode Gaussian initializations are considered. 
This happens on time intervals where the considered one-mode evolutions are incompatibility and entanglement breaking, nullifying Gaussian steerability and entanglement, respectively. 
Therefore, we considered a strategy that makes use of three-mode Gaussian states, namely the GHZ/W three-mode squeezed states, where Alice owns the first two modes which allows to overcome the problem of Gaussian incompatibility breaking and entanglement breaking.
For classical noise evolutions, we also showed that our use of the GHZ/W three-mode squeezed states allows to witness any non-Markovian evolution of this kind. 
However, in the more general case of evolutions characterized by lossy channels, 
our results show that, while the GHZ/W three-mode squeezed states allows witnessing non-Markovian evolutions that cannot be witnessed with two-mode initializations, this three-mode state could not be enough for detecting all non-Markovian dynamics. In order to increase the potential to witness non-Markovianity via correlation backflows, it would be interesting to consider three-mode initializations different from the GHZ/W states or to increase the size of the ancillas. 
Nonetheless, for the quantum Brownian motion, a particular instance of the lossy channels of experimental interest (but still more sophisticated than the classical noise channels), 
the 
GHZ/W three-mode squeezed state
strategy accomplishes the task of detecting non-Markovianity through correlation backflows for values of the parameters where any two-mode initialization fails. 

An interesting open question is to identify a mode geometry, and corresponding state, able to display a correlation backflow for all non-Markovian evolutions. Our results have shown that two modes are not enough for Gaussian steerability and entanglement, while three modes do provide an improvement. Are three modes enough? If not, is a finite number of modes enough? Note that in the finite dimensional case, solutions to this question were derived in~\cite{DDSshort,DDSMJ},  and~\cite{Janek}. 
In the first work, a correlation measure $C$ was introduced based on state distinguishability and it was shown, for an evolution acting on a system $A_1$ of dimension $d$, how to construct an initial state defined on systems $A_1A_2 B$ with respective dimensions $d$, $d+1$, and $2$, that displays a backflow in the correlation measure $C$ along the bipartition $A_1A_2|B$ for all invertible non-Markovian evolutions. 
This approach was generalized in~\cite{DDSMJ}, where the authors considered a slightly more complex arrangement that made use of larger ancillas $A_2$ and $B$ to prove that for all non-Markovian evolutions there always exist a correlation and an initial state that provide a backflow.
In~\cite{Janek}, again for an evolution acting on a system $S$ of dimension $d$, it was provided an initial state consisting of systems $A_1A_2A_3B$ with respective dimensions $d$, $d+1$, $2$, and $2$ such that an entanglement negativity backflow along the bipartition $A_1A_2A_3|B$ could be detected for all invertible non-Markovian evolutions. 

Whether a similar arrangement is possible in the Gaussian case remains an open question.
Nonetheless, as discussed in Sec.  \ref{sec:NMwitnesses}, we expect
that any invertible $n$-mode non-Markovian Gaussian evolution on $A_1$ can be witnessed with a correlation backflow along the bipartition $A_1A_2|B$, where $A_2$ and $B$ are respectively (at most) two  and one mode ancillary Gaussian systems, no matter the number $n$ of evolving modes. It would be interesting to start this study by checking whether it is always possible to obtain correlation backflows along the bipartition $A_1A_2|B$ when a generic one-mode non-Markovian Gaussian evolution is applied on $A_1$ and $A_2$ ($B$) is a two-mode (one-mode) ancilla.

\section{Acknowledgements} 
The authors would like to thank Saleh Rahimi-Keshari for illuminating discussions. This work is supported by the Spanish Government (FIS2020-TRANQI and Severo Ochoa CEX2019-000910-S), the ERC AdG CERQUTE, the AXA Chair in Quantum Information Science, Fundacio Cellex, Fundacio Mir-Puig and Generalitat de Catalunya (CERCA, AGAUR SGR 1381) and the Swiss National Science Foundation NCCR SwissMAP.

\bibliographystyle{ieeetr} 
\bibliography{CVbib}  

\appendix
\onecolumngrid
\section{CP-divisibility criterion for smooth Gaussian dynamics}
In the main text we introduced non-Markovianity as violation of CP-divisibility of the intermediate map. For a smooth dynamics, CP-divisibility is equivalent to having an infinitesimally CP-divisible dynamics. In other words, at any time $t$ and for infinitesimally small $\epsilon > 0 $ we should have
\begin{align}
    (T_{t+\epsilon},N_{t+\epsilon}) = (T_{t,\epsilon},N_{t,\epsilon})\circ (T_{t},N_{t}),
\end{align}
where the intermediate map $(T_{t,\epsilon},N_{t,\epsilon})$ is CP. On the one hand, since $\epsilon$ is small we expect the map to take the form 
\begin{align}
    (T_{t+\epsilon},N_{t+\epsilon}) = \left(T_{t} + \epsilon {\dot T}_t + {\cal O} (\epsilon^2),N_{t}+\epsilon {\dot N}_t + {\cal O} (\epsilon^2)\right),
\end{align}
which means any covariance matrix $\sigma_0$ evolves as follows
\begin{align}\label{eq:CP_inf_dir}
    \sigma_{t+\epsilon} =  (T_{t+\epsilon},N_{t+\epsilon}) \sigma_0 = T_t \sigma_0 T_t^T + Y_t + \epsilon \left(
    {\dot T}_t \sigma_0 T_t^T + {\dot T}_t \sigma_0 {\dot T}_t^T + Y_t^{\prime} 
    \right) + {\cal O}(\epsilon^2).
\end{align}
On the other hand, we should have
\begin{align}\label{eq:CP_inf_int}
    \sigma_{t+\epsilon} & = (T_{t,\epsilon},N_{t,\epsilon}) \sigma_t  = (T_{t,\epsilon},N_{t,\epsilon})\circ (T_{t},N_{t}) \sigma_0 \nonumber\\
    &= T_{t,\epsilon} T_{t} \sigma_0 T_{t}^T T_{t,\epsilon}^T + T_{t,\epsilon} N_t T_{t,\epsilon}^T + N_{t,\epsilon}.
\end{align}
By comparing Eqs.~\eqref{eq:CP_inf_dir} and \eqref{eq:CP_inf_int} we should have the following two equivalences
\begin{align}
    T_{t,\epsilon} T_{t} \sigma_0 T_{t}^T T_{t,\epsilon}^T & \equiv T_t \sigma_0 T_t^T + \epsilon \left(
    {\dot T}_t \sigma_0 T_t^T + T_t \sigma_0 {\dot T}_t^T + {\dot N}_t 
    \right) + {\cal O}(\epsilon^2)\nonumber\\
    \Rightarrow & T_{t,\epsilon} = I + \epsilon {\dot T}_t T_t^{-1} + {\cal O}(\epsilon^2)\label{eq:inf_1}\\
    T_{t,\epsilon} N_t T_{t,\epsilon}^T + N_{t,\epsilon} & \equiv N_t + \epsilon {\dot N}_t + {\cal O}(\epsilon^2) \nonumber \\
    \Rightarrow & N_{t,\epsilon} = \epsilon \left( 
    {\dot N}_t - {\dot T}_t T_t^{-1} N_t - N_t T_t^{-T} {\dot T}_t^T
    \right) + {\cal O}(\epsilon^2), \label{eq:inf_2}
\end{align}
where we use the convention $A^{-T} = (A^{T})^{-1} = (A^{-1})^T$ for any matrix $A$.
Furthermore, the positivity of the intermediate map implies
\begin{align}
    N_{t,\epsilon} + i\Omega - iT_{t,\epsilon} \Omega T_{t,\epsilon}^T \geqslant 0,
\end{align}
which by using Eqs.~\eqref{eq:inf_1} and \eqref{eq:inf_2}---and keeping the leading order in $\epsilon$---is equivalent to the following criterion
\begin{align}\label{eq:NM_criterion_inf}
    N_{t,\epsilon} + i\Omega - \left(i\Omega + \epsilon \left( {\dot T}_tT_t^{-1} {i}\Omega + {i}\Omega T_t^{-T}{\dot T}_t^T\right)   \right)& \geqslant 0 \nonumber\\
    \Rightarrow {\dot N}_t - \left(  {\dot T}_tT_t^{-1} ({i}\Omega + N_t) + ({i}\Omega+N_t) T_t^{-T}{\dot T}_t^T
    \right)\geqslant 0.
\end{align}
The above condition is a necessary and sufficient condition for the intermediate map being non-Markovian, i.e., if it holds at all times, the dynamics is Markovian, otherwise, it is not.
{
Furthermore, it can be deduced directly from Eq.~(8) of Ref.~\cite{illuminati2015}, substituting in that equation at first order in $\epsilon$, 
$X(t+\epsilon, 0)\rightarrow\dot{X}(t,0) \epsilon+{X}(t,0)$
and
$Y(t+\epsilon, 0)\rightarrow\dot{Y}(t,0) \epsilon+{Y}(t,0)$, where $X(t, 0)$ and $Y(t, 0)$ of Ref.~\cite{illuminati2015} are the matrices $T_t$ and $N_t$, respectively.
}
\section{Classical noise channel: potential of the GHZ/W three-mode squeezed state setup}\label{appendix:classical}
Here we show that using the initialization $\sigma_{3,r}$ \eqref{EPR}
one can witness any non-Markovianity of the classical noise evolution \eqref{toy-channel} by means of steering backflow. Interestingly, this happens for any value of the squeezing parameter $r$.
The impossibility of achieving the same result through the two-mode initialization $\sigma_{2,r}$ \eqref{sigma0-2modes} is also shown analytically.
Finally, we conclude the section with an explanatory example of oscillating noise.
\subsection{Three-mode scenario}
\label{3modes4anytoy}
Consider the state given by $\sigma_{3,r}$ with only the first mode undergoing the classical noise channel.
The total map is given by the pair $({I^{(3)}},\eta_t { I^{(1)}} \oplus 0^{(2)})$, where we keep using the notation where  $I^{(n)}$ ($0^{(n)}$) is the $2n\times 2n$ identity (null) matrix. 
The covariance matrix after the channel reads

\begin{align}
	\sigma_{3,r}(t) = \left(
	\begin{array}{ccc}
		\sigma(r) + \eta_t{I}^{(1)} & \epsilon(r) & \epsilon(r)\\
		\epsilon(r) & \sigma(r) & \epsilon(r)\\
		\epsilon(r) & \epsilon(r) & \sigma(r)
	\end{array}
	\right).
\end{align}
Whenever $\partial_t \eta_t \leqslant 0$ the dynamics is non-Markovian (see Eq.~\eqref{noiseint} and related discussion). We want to see it via steering,
in particular we use the first two modes (including the noisy one) in order to steer the third mode.
According to \eqref{ineqwiseman} the state is steerable from the first two modes to the third one if and only if the following is violated
\begin{align}
	\sigma_{3,r}(t) + i{ 0}^{(2)} \oplus \Omega^{(1)} \geqslant 0,
\end{align}
which is equivalent to the non-physicality of the covariance matrix
\begin{align}\label{Mappendix}
     {M}_B =   B -   C^T   A^{-1}   C \, ,
\end{align}
with ${ A}\equiv {\sigma_{3,r}(t)}(1:4,1:4)$, 
${ C}\equiv {\sigma_{3,r}(t)}(1:4,5:6)$, 
and ${ B} \equiv {\sigma_{3,r}(t)}(5:6,5:6)$---note that Eq. (\ref{Schur}) corresponds to Eq. (\ref{Mappendix}) for $\sigma=\sigma_{3,r}$.
This gives the matrix 
\begin{align}
   {M_B} =  \left(\begin{array}{cc}
\frac{\eta_t+3\,{\mathrm{e}}^{2\,r} +2\eta_t{\rm e}^{4r} }{2\,{\mathrm{e}}^{4\,r} +2\,\eta_t\,{\mathrm{e}}^{2\,r} +\eta_t\,{\mathrm{e}}^{6\,r} +1} & 0\\
0 & \frac{{\mathrm{e}}^{2\,r} \,{\left(2\,\eta_t+3\,{\mathrm{e}}^{2\,r} +\eta_t\,{\mathrm{e}}^{4\,r} \right)}}{\eta_t+2\,{\mathrm{e}}^{2\,r} +{\mathrm{e}}^{6\,r} +2\eta_t{\rm e}^{4r} }
\end{array}\right).
\end{align}
The symplectic eigenvalue of the matrix above, 
 \begin{equation}
 \label{nu-3modes-classical}
\nu_-=
\sqrt{\frac{e^{2r} \left[\left(e^{4r}+2\right) \eta_t+3 e^{2r}\right] [e^{2r} (2 e^{2r} \eta_t+3)+\eta_t]}{\left[\left(e^{4r}+2\right) e^{2r} \eta_t+2 e^{4r}+1\right] 
\left[e^{2r} \left(e^{4r}+2 e^{2r} \eta_t +2\right)+\eta_t \right]}}
~,
\end{equation}
is always lower than 1, since, considering the expression inside the square root, the denominator exceeds the numerator by the positive quantity
$(-1 + e^{4r})^2 [\eta_t + e^{2r} (2 + e^{2r} \eta_t)]$ and both the numerator and the denominator are positive.
Furthermore, the derivative with respect to $\eta_t$, 
\begin{equation}
\frac{\partial \nu_-}{\partial \eta_t}=
\frac{e^{2r} \left(e^{4r}-1\right)^2 \left[4 e^{2r} \left(2 e^{8r}+5 e^{4r}+2\right) \eta_t+9 \left(e^{8r}+e^{4r}\right)+\left(2 e^{12r}+7 e^{8r}+7 e^{4r}+2\right) \eta_t^2\right]}{2 \left[
\left(e^{4r}+2\right) e^{2r} \eta_t+2 e^{4r}+1\right]^2 
\sqrt{\frac{e^{2r} \left[\left(e^{4r}+2\right) \eta_t+3 e^{2r}\right] [e^{2r} (2 e^{2r} \eta_t+3)+\eta_t]}{\left[\left(e^{4r}+2\right) e^{2r} \eta_t+2 e^{4r}+1\right] \left[e^{2r} \left(e^{4r}+2 e^{2r} \eta_t+2\right)+\eta_t\right]}} \left[e^{2r} \left(e^{4r}+2 e^{2r} \eta_t+2\right)+\eta_t\right]^2}
,
\end{equation}
is also positive.
This shows that for all values of $r$ and $\eta_t$, 
one can always detect non-Markovianity.
 In particular, 
 in the cases of small and large squeezing parameter $r$, 
 the Gaussian steerability function is less cumbersome.

The symplectic eigenvalue around $r=0^+$, i.e.,  for epsilon-entangled states, reads
\begin{align}
    \nu_- = 1 - \frac{16}{9(\eta_t+1)}r^2 + {\cal O}(r^3).
\end{align}
which is always smaller than one. 
Gaussian steerability from Alice, who owns the first two modes, to Bob, who owns the third mode, reads
\begin{align}
    {\cal G}_{A\to B}(\sigma_{3,r}(t)) = -\log (\nu_-) \approx \frac{16}{9(\eta_t + 1)}r^2 + {\cal O}(r^3),
\end{align}
which is a monotonically decreasing function of $\eta_t$. Hence, whenever the dynamics is non-Markovian, i.e.,  $\partial_t \eta_t\leqslant 0$, the Gaussian steerability increases and one can detect non-Markovianity.
For large squeezing parameter $r$, instead, we get
\begin{align}
    {\cal G}_{A\to B}(\sigma_{3,r}(t)) = \frac{1}{2} \ln \left(\frac{e^{2 r}}{2 \eta_t}\right)+
    O(e^{-2 r}),
\end{align}
leading to the same conclusion as before.
These results show that the state remains always steerable from the first two modes to the third mode, no matter how large $\eta_t$ is.\\

\subsection{Two-mode scenario}
\label{sec-2modes-classical}
For a scenario with two modes and input $\sigma_{2,r}$ (Eq.~\eqref{sigma0-2modes}) one can find that the matrix $M_B$ (see Eq. (\ref{Schur})) reads
\begin{align}
    \left(\begin{array}{cc}
\frac{\eta_t\,\cosh \left(2\,r\right)+1}{\eta_t+\cosh \left(2\,r\right)} & 0\\
0 & \frac{\eta_t\,\cosh \left(2\,r\right)+1}{\eta_t+\cosh \left(2\,r\right)}
\end{array}\right).
\end{align}
{
We have 
\begin{align}
\label{nu-2modes-classical}
\nu_-=\frac{\eta_t\,\cosh \left(2\,r\right)+1}{\eta_t+\cosh \left(2\,r\right)}
~,
\end{align}
which is smaller than $ 1$ iff $\eta_t< 1$,
and
\begin{align}
\frac{\partial \nu_-}{\partial \eta_t}=
\frac{\cosh^2(2r)-1}{(\eta_t+\cosh(2r))^2}
\end{align}
is always positive.
This implies that for any $r> 0$,
the two-mode initialization setup cannot detect non-Markovianity if $\eta_t\geqslant1$, but it does if $\eta_t<1$ 
since the function ${\cal G}_{A\to B}(\sigma_{2,r}(t))=\max \{0,-\ln (\nu_-)\}$ is monotonically decreasing with $\eta_t$.
}

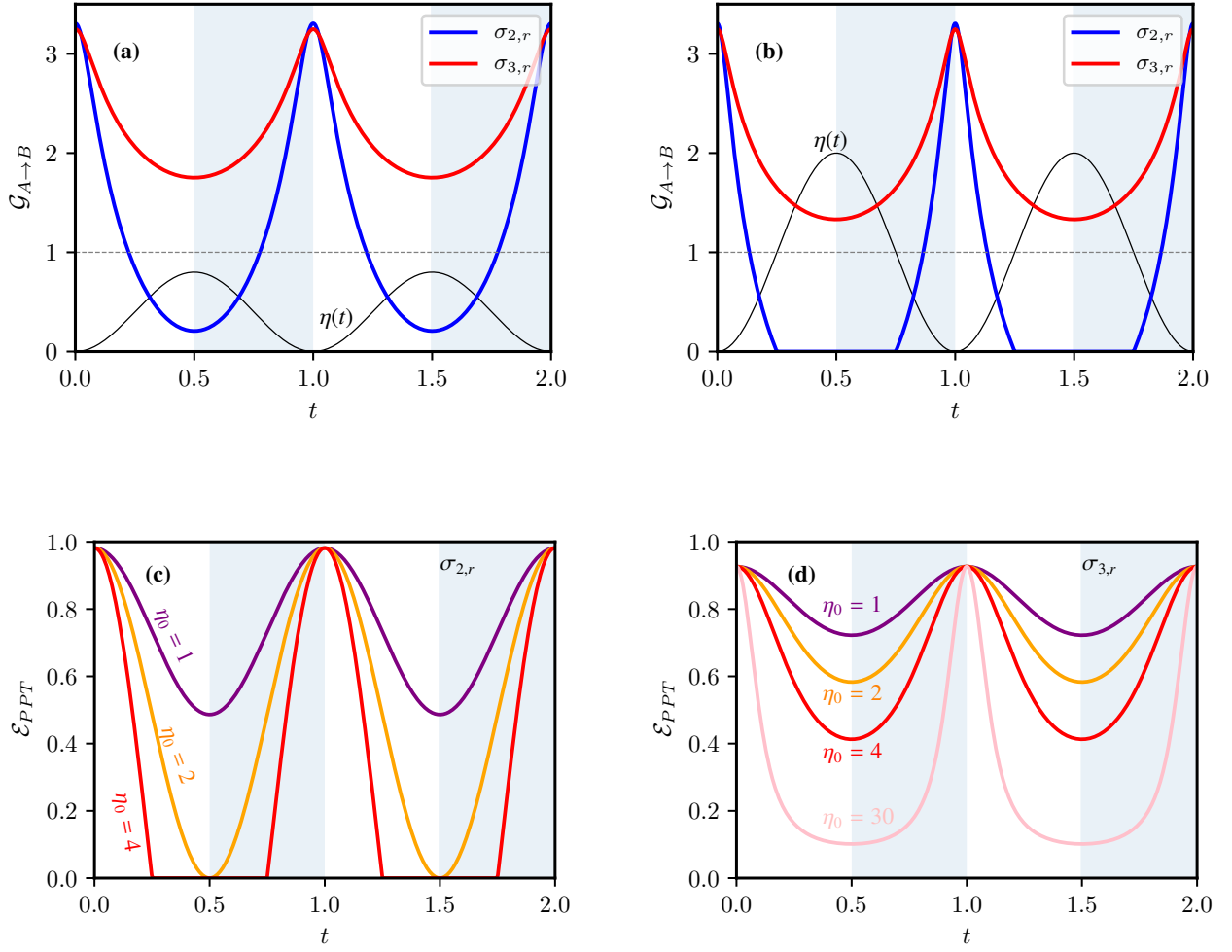
\begin{figure*}
\centering
\begin{overpic}[width=.45\linewidth]{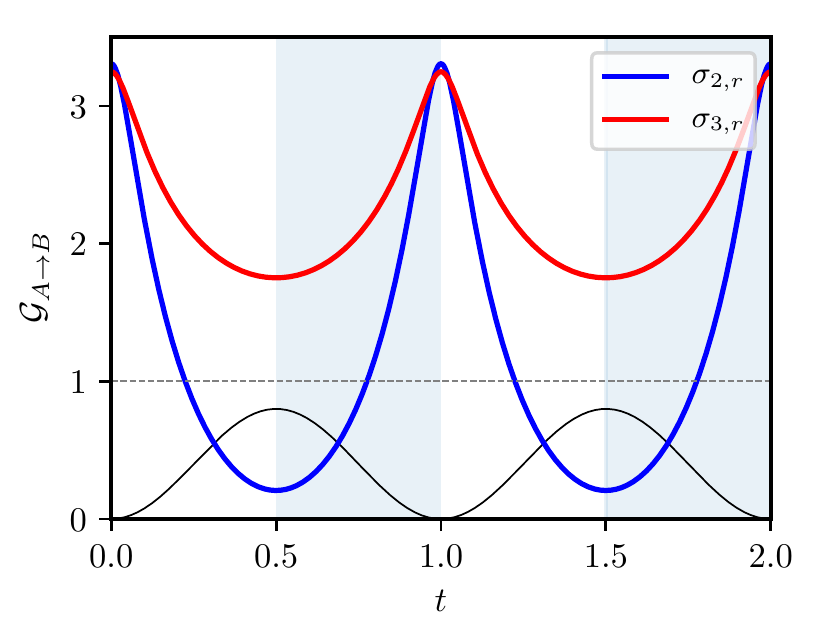}
\put(55,20){$\eta(t)$}
\put(20,65){\bf (a)}
\end{overpic}
\qquad
\begin{overpic}[width=.45\linewidth]{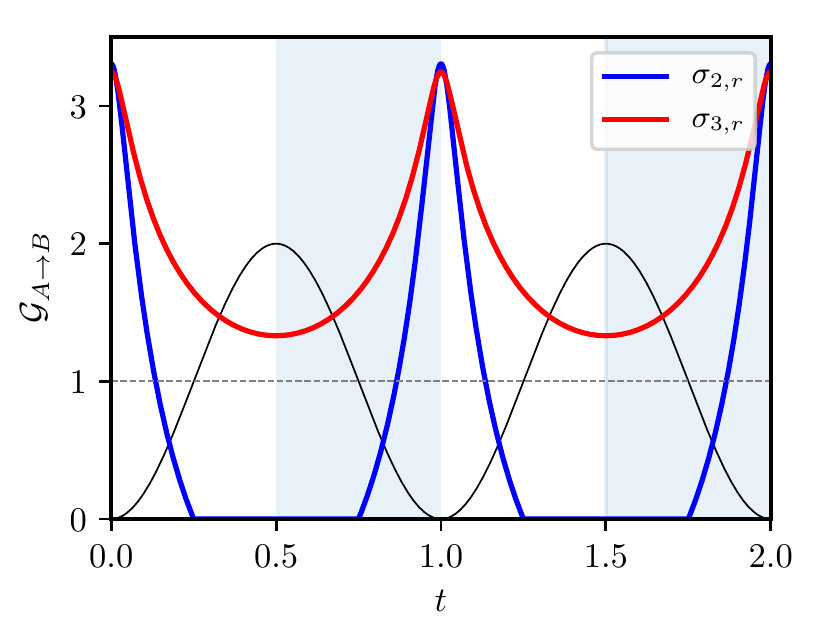}
\put(20,65){\bf (b)}
\put(30,50){$\eta(t)$}
\end{overpic}
\vspace{1cm}
\\
\begin{overpic}[width=.45\linewidth]{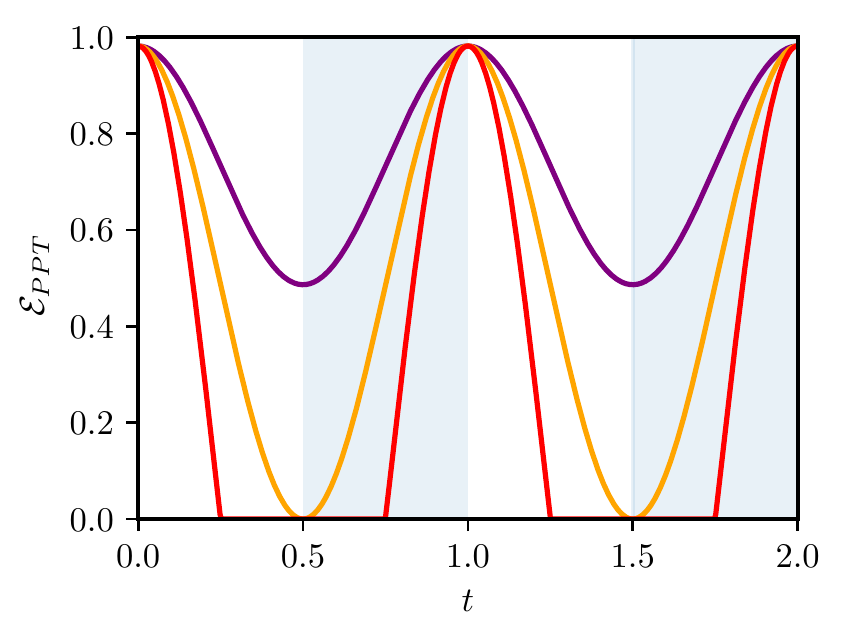}
\put(25,65){\bf (c)}
\put(25,60){   \color{violet} \rotatebox{-65}{$\eta_0=1$}}
\put(26,40){   \color{orange} \rotatebox{-70}{$\eta_0=2$}}
\put(18,30){   \color{red} \rotatebox{-80}{ $\eta_0=4$}}
\put(75,67){$\sigma_{2,r} $}
\end{overpic}
\qquad
\begin{overpic}[width=.45\linewidth]{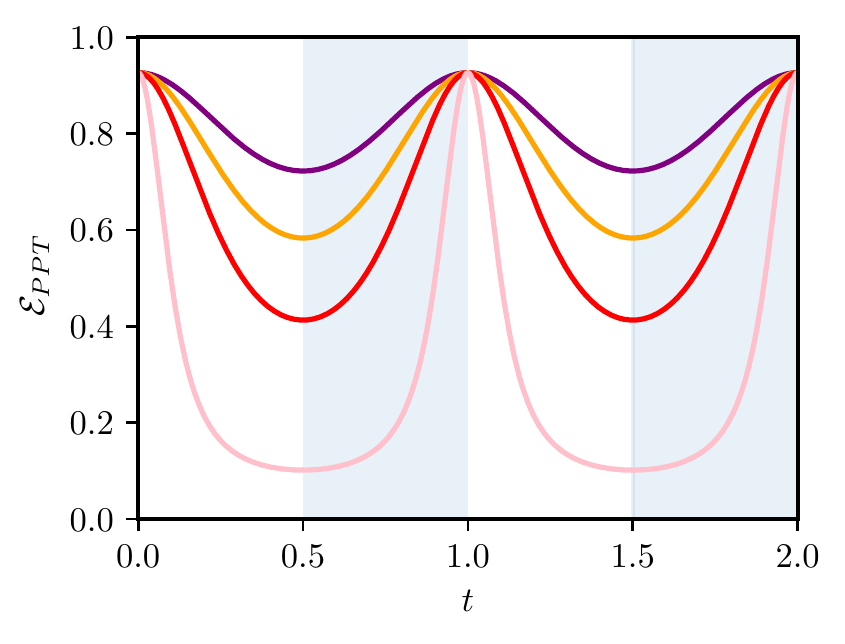}
\put(25,65){\bf (d)}
\put(30,60){   \color{violet} $\eta_0=1$}
\put(30,45){   \color{orange} $\eta_0=2$}
\put(30,35){   \color{red} $\eta_0=4$}
\put(30,24){   \color{pink} $\eta_0=30$}
\put(75,67){$\sigma_{3,r} $}
\end{overpic}
\\
\caption{
\label{FIGtoyOSCILLATING} (a)
Gaussian steerability $\mathcal{G}_{A\rightarrow B}(\sigma_{2,r}(t))$ (blue) and $\mathcal{G}_{A\rightarrow B}(\sigma_{3,r}(t))$ (red) with Alice's first mode evolving through the classical noise channel.
Setting
{$\eta(t)=\eta_0 (1-\cos(2\pi t))/2$ and $\eta_0=0.8$ (black), $\mathcal{G}_{A\rightarrow B}(\sigma_{2,r}(t))$ provides backflows whenever $\eta(t)$ is decreasing, namely for any non-Markovian behavior (blue shadow regions). In this case the dynamical maps are not GIB.}
(b)
Same as in (a) but for $\eta_0=2$: $\mathcal{G}_{A\rightarrow B}(\sigma_{2,r}(t))$ provides backflows whenever $\eta(t)$ is decreasing and is smaller than 1 (dashed line), but 
does not witness non-Markovianity when $\eta(t)$ is decreasing and above 1 (GIB threshold, dashed line).
(c)
Same as in (a) and (b), but for 
entanglement $\mathcal{E}_{PPT}(\sigma_{2,r}(t))$ 
and for values of $\eta_0$: 1 (purple), 2 (orange) and 4 (red). 
The dynamical maps of this evolution are periodically EB in finite time intervals for $\eta_0> 2$.
For this reason, some time intervals of non-Markovianity cannot be witnessed by backflows of entanglement. 
(d)
Same as in (c) but for the three-mode squeezed state  initialization
$\sigma_{3,r}(0)$
and for values of $\eta_0$: 1 (purple), 2 (orange), 4 (red) and 30 (pink). 
Since now Alice owns two modes but only the first is subjected to the noise, the dynamical maps of this evolution are never EB and all the non-Markovian behavior of the dynamics can be witnessed with entanglement backflows.
All the plots in this Figure are made for $r=2$.}
\end{figure*}

Again, behaviors for small and large squeezing parameter $r$ are the following. For small $r=0^+$, we have 
\begin{align}
    \nu_- = 1-2\frac{1-\eta_t}{1+\eta_t}r^2 + {\cal O}(r^3),
\end{align}
and therefore, at leading order, the Gaussian steerability from Alice, who owns the first mode, to Bob, who owns the second mode, reads
\begin{align}
    {\cal G}_{A\to B}(\sigma_{2,r}(t))=
    \begin{cases}
    2\frac{1-\eta_t}{1+\eta_t}r^2,~~& \eta_t < 1\\
    0,~~~~~~& \eta_t \geqslant 1
    \end{cases}
\end{align}
which{, consistently,} cannot detect non-Markovianity if $\eta_t\geqslant1$, but it does if $\eta_t<1$, since the function is monotonically decreasing with $\eta_t$.
{
For large $r$ we finally obtain
\begin{align}
    {\cal G}_{A\to B}(\sigma_{2,r}(t))=\max
    \left\{
    0, \ln\left(\frac{1}{\eta_t}\right)+{\cal O}(e^{-2 r})
    \right\}.
\end{align}
}
{
\subsection{Classical noise: oscillating noise}
\label{appendix:TOYoscillating}
We conclude the discussion on the classical noise channel of Sec.~\ref{sec:classical-noise} by considering an oscillating noise that didactically shows when the two-mode configuration \eqref{sigma0-2modes} fails in detecting non-Markovianity.
We hence set the noise as 
\begin{equation}
\eta(t)=\eta_0 (1-\cos(2\pi t))/2,
\end{equation}
with $\eta_0$ being the constant gauging its intensity.
If the constant $\eta_0 \leqslant 1$,
 an initialization in $\sigma_{2,r}(0)$ implies backflows of $\mathcal{G}_{A\rightarrow B}(\sigma_{2,r}(t))$ if and only if $\eta(t)$ decreases, see Fig. \ref{FIGtoyOSCILLATING}(a).
In this case two modes are enough to detect any non-Markovian character.
On the contrary, if $\eta_0>1$ we observe backflows of $\mathcal{G}_{A\rightarrow B}(\sigma_{2,r}(t))$ if and only if $\eta(t)$ decreases \textit{and} $\eta(t)<1$, see Fig. \ref{FIGtoyOSCILLATING}(b).
As analytically shown, the three-mode initialization $\sigma_{3,r}(0)$
does not suffer of this limitation providing steering backflow as soon as the map shows non-Markovian behavior and for any value of $\eta(t)$. 

Fig.~\ref{FIGtoyOSCILLATING}(c) and (d) concern instead
entanglement $\mathcal{E}_{PPT}$, 
again, for different values of $\eta_0$.
The core message does not change by considering this other correlation quantifier.
The dynamical maps of this evolution are periodically EB in finite time intervals for $\eta_0> 2$.
For this reason, whether some time intervals of non-Markovianity cannot be witnessed by backflows of entanglement in the two-mode configuration \eqref{sigma0-2modes} (see Fig.~\ref{FIGtoyOSCILLATING}(c)), the three-mode configuration
\eqref{EPR}
never fails in detecting non-Markovianity (see Fig.~\ref{FIGtoyOSCILLATING}(d)). 
}

\section{Lossy channel}
\label{appendix:lossy}
The lossy channel is given by $T_t=\tau_t I$ and $N_t = \eta_t I$, see Eq.~\eqref{lossy-channel}.
First of all, from Eqs.~\eqref{N1mode} and \eqref{CPTP1mode}, the complete positivity of the channel imposes 
\begin{equation}\label{cp-lossy-channel}
    \eta_t\geq \tau_t^2 \left\lvert 1-\frac{1}{\tau_t^{2}}\right\rvert ~,
\end{equation}
while $\tau_t$ can also be considered positive in full generality.
\subsection{CP-divisibility}
The CP-divisibility of the intermediate map breaks down if and only if Eq.~\eqref{eq:NM_criterion_inf} is violated.
% \begin{align}
%     {\dot N}_t - \left[ {\dot T}_t T_t^{-1} (\Omega + N_t) + (\Omega + N_t) T_t^{-T} {\dot T}_t^T
%     \right]\geqslant 0.
% \end{align}
For the lossy channel, the matrix on the l.h.s. of \eqref{eq:NM_criterion_inf} has the following eigenvalues
\begin{align}
    \lambda_{\pm} = {\dot \eta}_t - 2(\eta_t\pm 1)\frac{\dot \tau_t}{\tau_t}.
\end{align}
and therefore when the sign of at least one eigenvalue is negative we have non-Markovianity,
formally
\begin{align}\label{eq:lossy_cpdiv}
    \sign[\lambda_{\pm}] = \sign[\tau_t^2{\dot \eta_t} - 2\tau_t{\dot \tau_t}\eta_t \pm 2\tau_t{\dot \tau_t}],
\end{align}
getting the value $1$ iff we are in the Markovian case.
This proves inequality \eqref{eq:NM_LOSSY} of Sec.~\ref{BrownMot}.\\

\subsection{Gaussian steerability}
The two-mode Gaussian steerability, where Alice and Bob own one mode, reads 
\begin{align}
    {\cal G}_{A\to B}(\sigma_{2,r}(t))=\max\left\{0, \frac{2(\tau(t)^2-\eta(t))}{\tau(t)^2+\eta(t)}r^2 + {\cal O}(r^3)\right\}
\end{align}
for small $r$ and
\begin{align}
    {\cal G}_{A\to B}(\sigma_{2,r}(t))=\max
    \left\{
    0, \ln\left(\frac{\tau_t^2}{\eta_t}\right)+{\cal O}(e^{-2 r})
    \right\},
\end{align}
for large $r$.
As a consequence, one can check that,
both for the limits of small and large $r$,
\begin{align}
\label{2modes-lossy-steerBFappendix}
    {\sign}[\dot {{\cal G}}_{A\to B}(\sigma_{2,r}(t))] = \begin{cases}
    \sign[ -{\dot \eta(t)}+2\eta(t)\frac{{\dot \tau}(t)}{\tau(t)}], &\eta(t) < \tau^2(t)\\
    0, &\eta(t) \geqslant \tau^2(t)
    \end{cases}  \, .
\end{align}
{
This condition remains valid for any $r>0$.
Indeed,
we have 
\begin{align}
\label{nu-2modes-lossy}
\nu_-=\frac{(\eta_t/\tau_t^2)\,\cosh \left(2\,r\right)+1}{(\eta_t/\tau_t^2)+\cosh \left(2\,r\right)}
~,
\end{align}
i.e. nothing but Eq. \eqref{nu-2modes-classical} with the substitution 
$\eta_t\rightarrow \eta_t/\tau_t^2$. Hence,
\eqref{nu-2modes-lossy}
is smaller than $ 1$ iff $\eta_t< \tau_t^2$
and 
its time derivative is
$\dot{\nu}_-=\frac{\partial \nu_-}{\partial (\eta_t/\tau_t^2)} \frac{d (\eta_t/\tau_t^2)}{dt}$.
While $\frac{\partial \nu_-}{\partial (\eta_t/\tau_t^2)}$ is always positive (to see it one can directly use the results from Sec.~\ref{sec-2modes-classical}), $\sign[\frac{d (\eta_t/\tau_t^2)}{dt}]=\sign[\dot{\eta_t}-2 \eta_t \dot{\tau_t}/\tau_t]$.
This
proves condition \eqref{2modes-lossy-steerBFappendix} and \eqref{eq:GS_LOSSY_2m} for any $r>0$.}
\\

On the other hand, the three-mode Gaussian steerability (Alice owns the first two modes and Bob owns the third mode) for small $r$ and large $r$ reads, respectively, as
\begin{align}
    {\cal G}_{A\to B}(\sigma_{3,r}(t)) = \frac{16 \tau_t^2}{9(\tau_t^2+\eta_t)}r^2 + {\cal O}(r^3),
\end{align}
\begin{align}
    {\cal G}_{A\to B}(\sigma_{3,r}(t)) = \frac{1}{2}\ln\left( \frac{\tau_t^2 e^{2r}}{2 \eta_t}\right)+ {\cal O}(e^{-2r}),
\end{align}
which is always \textit{positive}. Therefore, Alice can (somewhat obviously) always steer the third mode.
One can easily check that,
both for the limits of small and large $r$,
\begin{align}
\label{3modes-steerBF-cond}
    {\sign}[{\dot {\cal G}}_{A\to B}(\sigma_{3,r}(t))] = \sign[ -{\dot \eta(t)}+2\eta(t)\frac{{\dot \tau}(t)}{\tau(t)}],
\end{align}
and therefore, one can detect non-Markovianity if the sign is positive, proving condition \eqref{eq:GS_LOSSY_3m}. However, it is a stricter constraint than \eqref{eq:lossy_cpdiv}, and thus we may not detect some non-Markovian dynamics.
{One can finally analytically check that condition \eqref{3modes-steerBF-cond} is valid for any $r>0$.
In this case, analogously to what we observed for expression \eqref{nu-2modes-lossy}, 
$\nu_-$ is nothing but expression \eqref{nu-3modes-classical} with the substitution 
$\eta_t\rightarrow \eta_t/\tau_t^2$.
Therefore, $\nu_-<1$ and the $\sign[\dot{\nu}_-]=\sign[ \dot{\eta}_t-2 \eta_t \dot{\tau}_t/\tau_t]$, proving \eqref{3modes-steerBF-cond} for any $r>0$.
}
\section{Quantum Brownian motion}
\label{appendix:meQBM}
The quantum Brownian motion is a particular example of the lossy channel \eqref{lossy-channel} admitting a microscopic derivation.
\subsection{Microscopic derivation and master equation}
We consider the following Hamiltonian $H$ for the whole system-environment compound, 
\begin{eqnarray}
&\hat{H}=\hat{H}_S+\hat{H}_E+\hat{H}_I~,&\\
&\hat{H}_S=\omega_0 \hat{a}^\dag \hat{a}~,\quad
\hat{H}_E=\sum_k \omega_k \hat{b}_k^\dag \hat{b}_k~,\quad
\hat{H}_I= \frac{\alpha}{2} \hat{q} \hat{Q},\quad \hat{Q}=\sum_k g_k \hat{Q}_k,&
\end{eqnarray}
where $\hat{H}_S$, $\hat{H}_E$ are the local terms on the system and on the environment, respectively,
$\hat{H}_I$ is the system-environment interaction term and $\hat{q}=\hat{a}+\hat{a}^\dag$ and $\hat{Q}_k=\hat{b}_k+\hat{b}_k^\dag$ are the position operators of the system (bosonic ladder operators $\hat{a}, \hat{a}^\dag$) and environment (bosonic ladder operators $\hat{b}_k, \hat{b}_k^\dag$).
$\hat{H}_I$ is a dipole-like interaction having coupling constant $\alpha$ controlling its strength.
Performing a second order expansion on the exact dynamics in interaction picture and enforcing Born (weak-coupling) and first Markov approximations one arrives to the Redfield equation for the reduced system dynamics in interaction picture 
\cite{book_B&P}
\begin{eqnarray}
\frac{d}{dt}\hat{\rho}(t)=
-\frac{\alpha^2}{4}\int_0^td \tau [\tilde{\hat{q}}(t) \tilde{\hat{q}}(t-\tau) \hat{\rho}(t)-
%\nonumber \\
\tilde{\hat{q}}(t-\tau)\hat{\rho}(t)\tilde{\hat{q}}(t)]
\langle 
\tilde{\hat{Q}}(\tau) \hat{Q}
\rangle_{T}+H.c.~,
\end{eqnarray}
where
$\langle \dots \rangle_{T} $
denotes the average over the bath thermal state at temperature $T$ and we indicated operators in interaction picture as 
$\tilde{\hat{\theta}}(t)= \exp[i (\hat{H}_S+\hat{H}_E) t] \hat{\theta} \exp[-i (\hat{H}_S+\hat{H}_E) t]$.\\
In the secular approximation, i.e.,  canceling the fast oscillating counter-rotating terms $\exp(\pm i 2 \omega_0 t)$, neglecting the Lamb shift and assuming the environment to be in the thermal state with temperature $T$, the above equation reduces to the quantum Brownian motion master equation (see e.g. \cite{maniscalco2004, ParisCV})
\begin{eqnarray}&&
\frac{d}{dt}\hat{\rho}(t)= \frac{\Delta(t)+\gamma(t)}{2} (\hat{a} \hat{\rho}(t) \hat{a}^\dag-\frac{1}{2}\{\hat{a}^\dag \hat{a}, \hat{\rho}(t) \})
+
%\nonumber \\
 %&&
% \qquad
 \frac{\Delta(t)-\gamma(t)}{2} (\hat{a}^\dag \hat{\rho}(t) \hat{a}-\frac{1}{2}\{\hat{a} \hat{a}^\dag, \hat{\rho}(t)\})
\end{eqnarray}
with $\Delta(t)$ and $\gamma(t)$ being the diffusion and damping coefficients, respectively, defined in \eqref{eq:Delta-and-gamma}
 and $J(\omega)=\sum_k g^2_k \delta(\omega-\omega_k)$
 is the spectral density (in the main text, assumed to be of the form \eqref{eq:specdens}).
This implies the following master equation for the $(1+n)$-mode covariance matrix 
\begin{eqnarray}
\label{me-sigma}
\frac{d}{dt}\sigma(t)=A(t) \sigma(t) + \sigma(t) A^T(t) +D(t)~,
\end{eqnarray}
with 
\begin{eqnarray}
A(t)= [-\frac{\gamma(t)}{2}  {I}^{(1)}] \oplus  { 0}^{(n)}~,\\
D(t)= [\Delta(t)  { I}^{(1)} ]  \oplus   { 0}^{(n)} ~,
\end{eqnarray}
where we assumed only the first mode to be affected from the channel.
In order to get the dynamics of second order moments we solved numerically the integrals \eqref{eq:Delta-and-gamma} and the master equation \eqref{me-sigma} for the covariance matrix. An equivalent approach would be to consider the integrated lossy channel expression (see, e.g., \cite{ParisCV}) 
\begin{eqnarray}
&&\sigma(t)=[(\tau(t)  { I}^{(1)}\oplus {I}^{(n)}]\sigma_0[(\tau(t)  {I}^{(1)})\oplus  {I}^{(n)}]^T+
%\nonumber\\
%&&
\eta(t) { I}^{(1)}\oplus  {  0}^{(n)}~,
\end{eqnarray}
with $\tau(t) $ and $\eta(t)$ given in \eqref{eq:Delta-and-gamma}.
\begin{figure*}
\centering
\begin{overpic}[width=.45\linewidth]{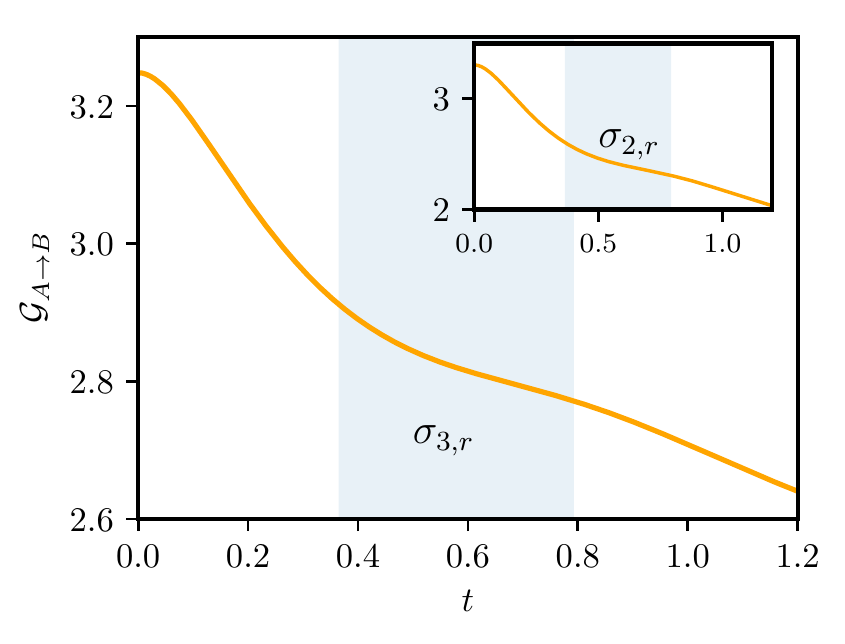}
\put(25,65){\bf (a)}
\end{overpic}
\qquad
\begin{overpic}[width=.45\linewidth]{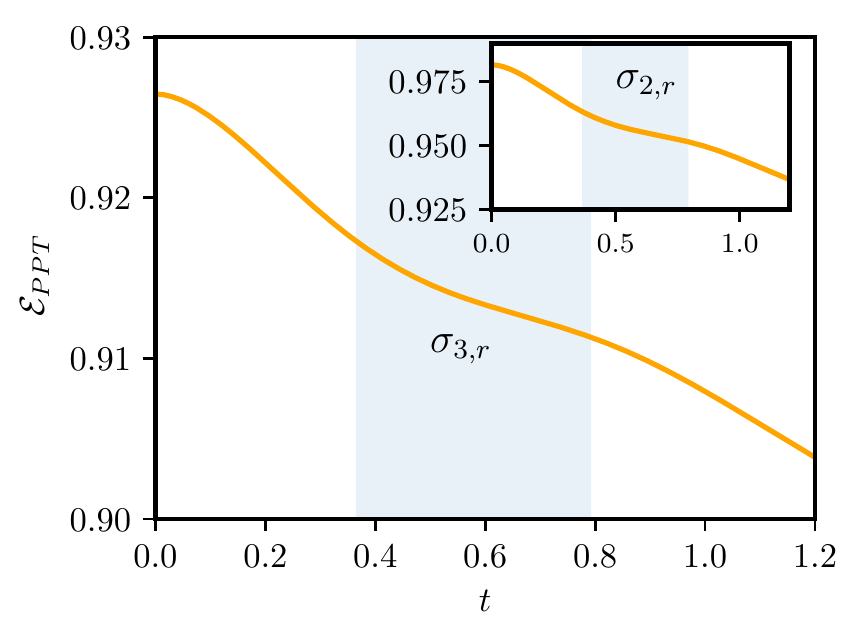}
\put(25,65){\bf (b)}
\end{overpic}
\caption{
\label{fig:QBM-low-temperature} %
(a)
Gaussian steerability $\mathcal{G}_{A\rightarrow B}$ as function of time for the three-mode squeezed state initialization $\sigma_{3,r}(0)$ and Alice's first mode subjected to the quantum Brownian motion,
for coupling parameter $\alpha=0.7$,
Ohmic regime ($s=1$), $r=2$, $\omega_0=7$, $\omega_c=1$, $T=0.5$ (low temperatures). 
Inset: $\mathcal{G}_{A\rightarrow B}$ for the two-mode squeezed state initialization $\sigma_{2,r}(0)$ with the same parameters. 
(b)
Same as in (a) but for 
entanglement $\mathcal{E}_{PPT}$.
Both $\mathcal{G}_{A\rightarrow B}$ and $\mathcal{E}_{PPT}$ do not show backflow in the non-Markovian region (shadow blue regions, identified by condition \eqref{eq:QBM-NM}), this happening for both the considered initializations. 
}
\end{figure*}

\subsection{Low temperature regime: failure of the GHZ/W three-mode squeezed state setup.}
\label{appendix:low-temperature}
Comparing the violation of the Markovian condition \eqref{eq:NM_LOSSY} with  the backflow conditions for Gaussian steerability \eqref{eq:GS_LOSSY_2m} and \eqref{eq:GS_LOSSY_3m}, we infer that any non-Markovianity can be detected in the limit $ \eta(t) \gg 1$ via steering backflow, at least for the three-mode initialization $\sigma_{3,r}(0)$. For quantum Brownian motion this limit is achieved at high temperatures, see Fig.~\ref{FIGbrownian}.
Nonetheless, by lowering the temperature one expects the limitation in sensitivity to manifest.
In Fig.~\ref{fig:QBM-low-temperature}(a),
we plot $\mathcal{G}_{A\rightarrow B}$ as a function of time for inputs $\sigma_{3,r}(0)$ and $\sigma_{2,r}(0)$ (inset) at low temperatures. 
We observe that $\mathcal{G}_{A\rightarrow B}$ does not show backflow in the non-Markovian region, for both the considered initializations.
In Fig.~\ref{fig:QBM-low-temperature}(b)
we also observe that the same insensitivity is obtained by considering entanglement $\mathcal{E}_{PPT}$.

\end{document}